\documentclass[10pt]{article}

\usepackage{amsmath}
\usepackage{graphicx,psfrag,epsf}
\usepackage{enumerate}
\usepackage{amsfonts}
\usepackage{amssymb}
\usepackage[usenames,dvipsnames]{xcolor}
\usepackage{tikz}
\usepackage{url}
\usetikzlibrary{arrows}
\usepackage{enumitem}
\usepackage{mathabx}
\usepackage{natbib}
\newtheorem{theorem}{Theorem}
\newtheorem{assumption}{Assumption}
\newcommand{\Esp}{\mathbb{E}}
\newcommand{\pen}{\text{pen}}
\providecommand{\keywords}[1]{\textbf{\textit{Keywords}} #1}

\begin{document}

\author{ Genevi\`{e}ve Robin$^{1}$, Christophe Ambroise$^{2}$ \\
and Sté\'ephane Robin$^{3}$\\
$^{1}$CMAP, UMR 7641, 
	  \'{E}cole Polytechnique,
	  X-POP, INRIA,\\	  
      Palaiseau,
      France\\
$^{2}$LaMME,
	  Universit\'e Paris-Saclay,
	  Universit\'e d'\'Evry val d'Essonne,\\
	  \'{E}vry,
	  France\\
$^{3}$ MIA-Paris, AgroParisTech,
	  INRA, Universit\'e Paris-Saclay,\\
	  Paris,
	  France
}
\title{Incomplete graphical model inference via latent tree aggregation}
\maketitle
\begin{abstract}
Graphical network inference is used in many fields such as genomics or ecology to infer the conditional independence structure between variables, from measurements of gene expression or species abundances for instance. In many practical cases, not all variables involved in the network have been observed, and the samples are actually drawn from a distribution where some variables have been marginalized out. This challenges the sparsity assumption commonly made in graphical model inference, since marginalization yields locally dense structures, even when the original network is sparse. 
We present a procedure for inferring Gaussian graphical models when some variables are unobserved, that accounts both for the influence of missing variables and the low density of the original network. Our model is based on the aggregation of spanning trees, and the estimation procedure on the Expectation-Maximization algorithm. We treat the graph structure and the unobserved nodes as missing variables and compute posterior probabilities of edge appearance. To provide a complete methodology, we also propose several model selection criteria to estimate the number of missing nodes. A simulation study and an illustration flow cytometry data reveal that our method has favorable edge detection properties compared to existing graph inference techniques. The methods are implemented in an R package.
\end{abstract}

\section{Introduction}

\subsection{Motivations}

Graphical models have been extensively studied and used in a wide variety of contexts, to represent complex dependency structures. 
In many practical cases however, it is more than likely that some variables involved in the network were in fact not observed.
Such missing variables are interpreted as actors that were not measured but nonetheless influence the measurements, or experimental conditions that were not taken into account. In the perspective of unrevealing the conditional independence structure, this can lead to both inference issues and interpretation problems.\\
The existence of unobserved variables can be naturally encompassed in the graphical model framework, by assuming there exists a 'full' graph describing the conditional independence structure of the joint distribution of observed and hidden variables. Observations are then samples of the marginal distribution of the observed variables only. From a graph-theoretical point of view, marginalizing hidden variables means removing them from the node set and marrying their children together, thus forming complete subgraphs, \textit{i.e.} cliques. Hence, the conditional independence structure among observed variables is described by a marginal graph containing locally dense structures. This violates the sparsity assumption on which the majority of graph inference methods are based. Moreover, an identifiability problem arises in the hidden variable setting, since infinitely many full graphs induce the same marginal structure.\\
In this paper we are interested in both checking if some variables are indeed missing in the graph and, if it is the case, inferring the complete graphical model. We address these problem in the context of Gaussian graphical models.

\subsection{Incomplete Gaussian graphical models}

Consider a multivariate Gaussian random vector parametrized by its precision matrix
\begin{equation}
\label{eq:x-gaus}
X \in \mathbb{R}^{p+r}\sim\mathcal{N}(0,K^{-1}),\qquad p,r\geq 1,\qquad K\in\mathbb{R}^{(p+r) \times (p+r)}\succ 0,
\end{equation}
where $\succ $ denotes positive definiteness. We assume that $X$ can be decomposed as 
$$
X = (X_O,X_H), 
$$
where $X_O \in \mathbb{R}^p$ denotes a set of observed variables and $X_H \in \mathbb{R}^r$ a set of hidden variables. In genomics, the hidden variables are understood as genes or experimental conditions that were not measured but nonetheless influence the results of the experiments. The goal of graphical model inference is to uncover the conditional independence structure of $X$, described by the following \textit{full graph}
\begin{equation}
\label{eq:full-graph}
G = (\{1,\dots ,p, p+1, \dots , p+r\}, E),
\end{equation}
where $E$ is the set of undirected edges, such that $\{i,j\}\in E$ if and only if $X_i$ and $X_j$ are dependent conditionally to $X_{\{1,\dots, p+r\}\setminus\{i,j\}}$, which we denote $X_i\not\perp X_j | X_{\{1,\dots, p+r\}\setminus\{i,j\}}$. In the Gaussian setting we consider, the set of edges $E$ is nicely determined by the non-zero entries of $K$ \citep{laurGM}: 
\begin{equation}
\label{eq:K-aretes}
\text{For all } (i,j)\in\{1,\ldots,p+r\}^2\text{, }i\neq j\text{, }\{i,j\}\in E \text{ if and only if } K_{ij}\neq 0.
\end{equation}
The precision matrix $K$ can be written block-wise to differentiate the terms corresponding to observed and latent variables:
\begin{equation} \label{eq:Kmatrix}
K = \begin{pmatrix}
K_O & K_{OH} \\ 
K_{HO} & K_H
\end{pmatrix}.
\end{equation}
From \eqref{eq:Kmatrix} and the Schur complement formula \cite[Example 3.15]{Boyd:2004:CO:993483} we deduce that the marginal distribution of the observed variables is
\begin{equation}
\label{schur}
X_O \sim \mathcal{N}(0, K_m^{-1}),\quad K_m = K_O - K_{OH}K_H^{-1}K_{HO}.
\end{equation}
The conditional independence structure of $X_0$ is thus described by the following \textit{marginal graph}
$$
G_m = (\{1,\dots ,p\}, E_m),
$$
where $E_m$ is the set of undirected edges given by the non-zero entries of $K_m$. Consider a sample $(X_O^{1},\ldots, X_O^{n})$ of $n$ independent realizations of the marginal distribution of $X_O \sim \mathcal{N}(0, K_m^{-1})$. From such measurements, standard statistical tasks are to infer the \textit{full graph} $G$ or the \textit{marginal graph} $G_m$; in this article we tackle both problems.

\subsection{Contributions and related work}
Methods to perform graphical model inference with unobserved variables have been proposed in the past. Some use the Expectation-Maximization (EM) algorithm \citep{DLR77}, its variational approximation described in \cite{VBEM}, or the Bayesian structural EM algorithm  \citep{BStructEM}. A lot of attention has also been brought to a regularized approach described in \cite{LatentCWP}, based on the analysis of the sum of low-rank and sparse matrices. Alternatives based on this method were also proposed by \cite{LLVGGM}, \cite{EMlvggm} and \cite{GirLatent}. \\
A major concern in the latent variable framework is identifiability; in general, identifiability constraints are very complex, as those derived in \cite{LatentCWP} for their model, which rely on algebraic geometry properties of low-rank and sparse matrices. On the contrary, in the particular case of trees (acyclic graphs), the conditions for identifying the joint graph from the marginal graph only, described in \cite{pearl}, are very simple. In this article, we propose to exploit this property to build an inference strategy based on the EM algorithm and spanning trees. \\
Latent tree models were studied in the context of phylogenetic tree learning; the Neighbor-Joining algorithm \citep{NJ} among others is a popular method in this field. More recently, a method called Recursive Grouping was proposed in \cite{LatentTreeGM}, to reconstruct tree structures from partially observable data. We emphasize the fact that all these methods learn a single tree from data. In the present, we take advantage of two key properties of tree-structured graphical models. First, we can specify under which conditions they remain identifiable in presence of missing variables. Second, treating trees as random, we can easily integrate over the whole set of spanning trees, thanks to an algebra result called the Matrix-Tree theorem \citep{Cha82}. To our knowledge, no method for latent variable graphical model inference is based on mixtures of trees, which constitute the main novelty of our approach. \\

Our contribution can be casted in the framework of \cite{MixtTrees}, who considered a special mixture of Bayesian network \citep[as defined by][]{GeH96} where each network involved in the mixture is tree-shaped. \cite{MixtTrees} show the interest of such a model both in terms of tractability and interpretation. \cite{MeJ06} also use the same framework to estimate the joint distribution of the observed variables and \cite{SZA15} aim at characterizing such distributions, but none of them is interested in the inference of the structure of the graphical model itself. A first difference with these tree-based methods is that we do not limit ourselves to a fixed number of trees but consider a mixture over all possible trees. Second, and more importantly, we extend the framework to the hidden variable setting.\\
Our inference strategy is based on the EM algorithm. The computations at the E step are tractable thanks to the Matrix-Tree theorem, which enables us to integrate over the whole set of spanning trees, as opposed to the M step of \cite{MixtTrees} that relies on the Chow-Liu algorithm \citep{ChowLiu}. This approach enables us to compute posterior probabilities of edge appearance, as proposed by \cite{BT} in the fully observable setting. To our knowledge, no other existing approach provides such an edge-specific measure of reliability. The final inference of the graph relies on the ranking of these probabilities, therefore we estimate graphs with general structures, though our method is based on trees. 
Although we mostly focus on the inference of the graph structure, we also obtain an estimate of the precision matrix of the joint distribution of the observed and hidden variables, as a by-product of the EM algorithm.\\

Our first contribution is to define, in Section \ref{model}, a latent tree aggregation model for graphical model inference in the presence of hidden variables {and to give identifiability conditions}. In Section \ref{inference}, we introduce our procedure based on the EM algorithm to infer the parameters of the joint distribution and probabilities of edge appearance, and to estimate the number of missing nodes. In Section \ref{experiments} we show on synthetic data that our method compares favorably to competitors in terms of edge detection. Finally we illustrate the procedure on flow cytometry data analysis in Section \ref{subsec:cytoData}.

\section{Latent Tree Aggregation Model}
\label{model}
\subsection{Identifiability conditions}
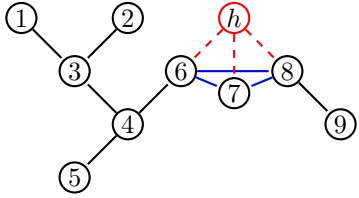
\begin{figure}[h]
\begin{center}
\begin{tikzpicture}[scale=0.3,->,>=stealth',shorten >=1pt,auto,node distance=1cm,
        thick,main node/.style={circle,draw,minimum size=0.4cm,inner sep=0pt]}]

    
%
        
    \node[main node] (1) at (-5,0) {$1$};
    \node[main node] (3) [below right of=1] {$3$};
    \node[main node] (2) [above right of=3] {$2$};
    \node[main node] (4) [below right of=3] {$4$};
    \node[main node] (5) [below left of=4]  {$5$};
    \node[main node] (6) [above right of=4] {$6$};
    \node[main node, draw=red] (10) [above right of=6] {$h$};
    \node[main node] (7) [below of=10] {$7$};
    \node[main node] (8) [below right of=10] {$8$};
    \node[main node] (9) [below right of=8] {$9$};
       
       \path[-]
    (1) edge node {} (3);
       \path[-]
    (2) edge node {} (3);

    		\path[-]
    	(4) edge node {} (5)
    		edge node {} (3)
		edge node {} (6);
		\path[-]
	(6) edge[blue] node {} (7)
		edge[blue] node {} (8);
		\path[-]
	(8) edge[blue] node {} (7)
		edge node {} (9);
		\path[-]
	(10) edge[red, dashed] node {} (6)
		edge[red, dashed] node {} (7)
		edge[red, dashed] node {} (8);
\end{tikzpicture}
\end{center}
\caption{Effect of marginalizing one hidden variable (h). Full  graph (all edges except blue), marginal graph (all edges except red).\label{fig:graph-types}}
\end{figure}

Assume the \textit{full graph} $G$ defined in \eqref{eq:full-graph} is tree structured. We now characterize the class of trees that are statistically identifiable in our model, \textit{i.e.} such that the full graph $G$ is uniquely determined by the marginal structure $G_m$. We assume without loss of generality that the observed and hidden variables are ordered, \textit{i.e.} $X_i$ is observed for all $i\in\{1,\ldots,p\}$ and hidden for all $i\in\{p+1,\ldots,p+r\}$, and denote for some set $A$ by $\operatorname{Card}(A)$ its cardinality. For $i\in \{1,\ldots p+r\}$, we define
$$
E_i = \left\lbrace j\in \{1,\ldots p+r\}\text{; }\{i,j\}\in E\right\rbrace .
$$
The following conditions on $G$ and $K$, derived from \cite{latentcause}, \cite{pearl} and \cite{LatentTreeGM}, guarantee statistical identifiability. 
\begin{assumption}[Identifiability conditions]\ \\
\vspace{-0.5cm}
\begin{enumerate}[label=(\roman*)]
\item For all $(i,j)\in\{p+1,p+r\}^2$, $\{i,j\} \notin E$;
\item For all $i\in\{p+1,p+r\}$, $\operatorname{Card}(E_i)\geq 3$;
\item Two nodes connected by an edge are neither perfectly independent nor perfectly dependent.
\end{enumerate}
\end{assumption}
These conditions stem from the simple graphical properties of spanning trees. Indeed, the maximal cliques of a tree are of size two, therefore if \textit{(i)} no edge connects two hidden nodes and \textit{(ii)} all hidden variables have at least three neighbors, there is exactly  one hidden node for every clique of size more than or equal to $3$ in $G_m$, as illustrated in Figure \ref{fig:graph-types}, and the class of identifiable trees is now fully characterized. In particular, hubs (central hidden nodes) are identifiable, while recovering chains of hidden nodes, or hidden nodes located at the leaves of the tree, is hopeless. An important feature is that our identifiability conditions allow sparsity in $G_m$, contrary to what happens in the sparse plus low-rank model of \cite{LatentCWP}. Indeed, identifiable graph structures in their case will typically have a small number of central hidden variables (hubs), and marginal graphs will therefore be densely connected, nay complete. This is an important difference with our model, and we will see in Section~\ref{experiments} that the inferred marginal structures are in fact very different.\\

\subsection{Fixed unknown tree} 
We now turn to the description of our Latent Tree Aggregation model, and start with a simple procedure where we infer a single tree structure. Let $\mathcal{T}$ be the set of spanning trees with $p+r$ nodes, and assume the graphical model associated with $X$, that we now write $T\in \mathcal{T}$, is tree-shaped. 
Assume further that, conditionally on $T$, the vector $X = (X_O, X_H)$ is drawn from the Gaussian distribution $\mathcal{N}(0, K_T^{-1})$, where $K_T$ has a tree-structured support determined by the edges of $T$, and can be decomposed in
\begin{equation} \label{eq:KTmatrix}
K_T = \begin{pmatrix}
K_{T,O} & K_{T,OH} \\ 
K_{T,HO} & K_{T,H}
\end{pmatrix}.
\end{equation}
In the complete data setting where $X$ is fully observed but $T$ is unknown, the Chow-Liu algorithm \citep{ChowLiu} computes the tree of maximum likelihood $\hat{T}$ from empirical observations, and the coefficients of the matrix $K_{\hat{T}}$ can be computed easily using a result of \cite{laurGM} and the empirical covariance matrix. Building $\hat{T}$ in this case boils down to finding a maximum spanning tree, which can be done with Kruskal's algorithm \citep{Kruskal}. If variables are now hidden but the underlying tree $T$ and $K_T$ are known, the conditional distribution of the hidden variables given the observed ones is $$X_H|X_O\sim\mathcal{N}(\mu_{H|O}, K_{H|O}^{-1}) ,\quad \mu_{H|O} = -K_{T,HO}X_O,\quad K_{H|O} = K_{T,H}.$$
From these two results, we can derive an EM algorithm to infer the tree-structured graph underlying the distribution of $X$ in the hidden variables setting, which runs iteratively until convergence, with the following steps at iteration $h+1$, $h\geq 1$.
\begin{description}
 \item[E-step:] Evaluation of the conditional expectation of the complete log-likelihood with respect to the current value $K^h$ of the parameter, namely:
 \begin{equation} \label{eq:CondExpCompLik1}
 \Esp_{X_H| X_O; K^h} \log p(X_O, X_H; K).
 \end{equation}
 \item[M-step:] Maximization of \eqref{eq:CondExpCompLik1} with respect to $K$ to update $K^h$ into $K^{h+1}$, using the Chow-Liu algorithm.
\end{description}

\subsection{Random unknown tree} 
The inference method described above is very simple, but the tree assumption is restrictive, and we expect poor results when it is violated. To overcome this, we choose to treat $T$ as a random variable. Doing so, we are able to compute a posterior probability of appearance for every possible edge in the graph. Ranking them in the decreasing order, we can infer a graph of general structure, even though our model is based on spanning trees. Denote by $E_T$ the set of edges of $T$. We assume $T$ to be drawn from a distribution defined by a matrix $\pi$ such that 
$$\pi_{ij} = P(\{i,j\}\in E_T).$$
The edges of $T$ are drawn independently, such that
\begin{equation} \label{eq:pT}
P(T)\propto \prod_{\{i,j\}\in E_T}\pi_{ij}. 
\end{equation}
Prior information about the existence of each edge is easily encoded in a distribution of this form, and a non-informative choice of prior is to set the $\pi_{ij}$ to be equal for all $i,j$, i.e. all trees have the same probability to be drawn so every edge has the same probability to be part of the drawn tree. 
We then assume the existence of a full symmetric matrix $K$ with block decomposition given in \eqref{eq:Kmatrix}, the entries of which have to be estimated.
For every $T\in \mathcal{T}$ we define the corresponding $(p+r)\times(p+r)$ matrix $K_T$, with off-diagonal term $K_{T, ij} = K_{ij}$ if $\{i,j\}\in E_T$ and zeros otherwise. The diagonal term $K_{T, ii}$ both depend on $K_{ii}$ and on the degree of node $i$ in $T$. Its expression derived from \cite{laurGM} is given in \eqref{Eq:defKT}, Appendix~\ref{appendix-1}. Note that $K$ does not need to be positive definite, although it may be desirable for the numerical stability of the algorithm. The joint distribution of $(X_O, X_H)$ is a mixture of centered Gaussian distributions:
$$
(X_O, X_H) \sim \sum_{T \in \mathcal{T}} p(T) \mathcal{N}(X_0, X_H; 0, K_T^{-1}).
$$
We develop this random unknown tree model further in Section \ref{inference} where we propose an inference procedure. For every possible edge $\{i,j\}$, we will compute the quantity
$$\alpha_{ij} = \sum_{\substack{T\in\mathcal{T}\\T\ni \{i,j\}}}P(T|X_O),$$
that we interpret as edge specific probabilities of appearance. First, we derive conditional distributions that will be necessary. In particular, we show that these distributions factorize over the edges.

\subsection{Some conditional distributions} 
Let us first compute the joint distribution of $T$ and $X_H$ conditionally on $X_O$ which will be needed in Section \ref{inference}:
$$P(T, X_H |X_O) = P(T |X_O)P(X_H|X_O, T).$$
On the one hand $P(X_H|X_O,T) =\mathcal{N}(\mu_{H|O,T},K_{H|O,T}).$ On the other hand, 
\begin{equation}
\begin{aligned} \label{eq:pTXo}
&P(T|X_O) &&\propto P(T)P(X_O|T)\\
& &&\propto\left(\prod_{\{i,j\}\in E_T} \pi_{ij}\right)\underbrace{\frac{\det(K_{T,m})^{\frac{n}{2}}}{(2\pi)^{\frac{np}{2}}}}_{(1)}\underbrace{\exp(-\frac{n}{2}\text{tr}(K_{T,m}\Sigma_O))}_{(2)},
\end{aligned}
\end{equation}
{where }
$K_{T,m}=K_{T,O}-K_{T,OH}(K_{T,H})^{-1}K_{T,HO}$. Terms (1) and (2) can be expressed as products over the edges of $T$. We directly give the results and leave the derivations to Appendix~\ref{appendix-1}. Let us define
\begin{equation}
\begin{aligned}
& d_{ij} && =\left(\frac{K_{ii}K_{jj}-K_{ij}^2}{K_{ii}K_{jj}}\right)^{\frac{n}{2}}\\
& t_{ij} && =\text{exp}\left(-nK_{ij}\Sigma_{ij}\right)
\end{aligned}
\quad \quad \forall \{i,j\}\in \{1,\ldots,p\}^2,
\end{equation} 
\begin{equation}
\begin{aligned}
f_{ih} = \text{exp}\left(\frac{n}{2}\sum_{k\in O}\frac{K_{ih}K_{hk}\Sigma_{ki}}{K_{hh}}  \right)\end{aligned}
\quad \quad \forall \{i,h\}\in \{1,\ldots,p\}\times \{p+1,\ldots,p+r\}
\end{equation} 
and finally 
\begin{equation}
\begin{aligned}
m_{ij} = \left\{
    \begin{array}{ll}
        t_{ij} & \mbox{if } \{i,j\}\in \{1,\ldots,p\}^2\\
        f_{ij} &\mbox{if } \{i,j\}\in \{1,\ldots,p\}\times \{p+1,\ldots,p+r\}\\
        f_{ij} &\mbox{if } \{i,j\}\in \{p+1,\ldots,p+r\}\times\{1,\ldots,p\} \\
        1 &\mbox{if } \{i,j\}\in \{p+1,\ldots,p+r\}^2\\
    \end{array}
\right..
\end{aligned}
\end{equation} 
We obtain that the conditional distribution  $P(T|X_O)$ nicely factorizes over the edges of $T$:
\begin{eqnarray}
P(T|X_O) & \propto & P(T) P(X_O|T) {\; \propto \prod_{\{i,j\}\in E_T} \pi_{ij} d_{ij} m_{ij}.} 
\end{eqnarray}

We also need to compute the normalizing constant of $P(T)$ and $P(T|X_O)$ -- that is, respectively,
$$
\sum_T \prod_{\{i, j\} \in E_T} \pi_{ij}
\quad \text{and} \quad
\sum_T \prod_{\{i, j\} \in E_T} \pi_{ij} d_{ij} m_{ij}.
$$

Those constants can be computed with the same complexity as a determinant, \textit{i.e.} in $O(p^3)$ operations, using the Matrix-Tree theorem that we now state. For a matrix $W$ of weights $w_{ij}$, we define the Laplacian
$\Delta = (\Delta_{ij})_{i,j\in V^2}$ associated to matrix $W$ by
$$\Delta_{ij}= \left\{\begin{matrix}
-w_{ij}\quad \quad \text{if }i\neq j, \\ 
\sum_{j}w_{ij}\quad \text{if } i=j.
\end{matrix}\right.$$
\begin{theorem}[\cite{Cha82}] \label{thm:MatTree}
Let $W=(w_{ij})_{(i,j)\in V^2}$ be a symmetric matrix of weights and $\Delta$ its associated Laplacian. For $(u,v)\in V^2$, let $\overline{\Delta}_{uv}$ be the $(u,v)$-th minor of $\Delta$. Then all $\overline{\Delta}_{uv}$ are equal and
$$
\overline{\Delta}_{uv} = \sum_{T\in \mathcal{T}}\prod_{\{i,j\}\in E_T}w_{ij} :=Z(W).
$$
\end{theorem}
In Section \ref{inference}, we will need to compute similar quantities after removing a given edge. Furthermore, we will need to compute such a quantity for all possible edges. This can be achieved in an efficient manner for all edges at a time thanks to a corollary of Theorem \ref{thm:MatTree} given in \cite{Kir07}, Theorem 3.
\section{Inference of the random unknown tree model} \label{inference}

\subsection{EM algorithm}

Because the proposed model involves unobserved variables, the EM algorithm \citep{DLR77} is a natural framework to carry the inference out. Importantly, two hidden layers appear in the model: the latent tree $T$ and the signal at the unobserved nodes $X_H$. We show that these two hidden layers can be handled, thanks to the matrix-tree theorem \citep{Cha82} introduced in Section \ref{model}. We first remind that the EM algorithm aims at maximizing the log-likelihood of the observed data $\log p(X_O; K)$ with respect to the parameter $K$, alternating two steps in an iterative manner. At iteration $h$ we perform:
\begin{description}
 \item[E-step:] Evaluation of all the conditional moments involved in the the conditional expectation of the complete log-likelihood with the current value $K^h$ of the parameter, namely:
 \begin{equation} \label{eq:CondExpCompLik}
 \Esp_{X_H, T | X_O; K^h} \log p(X_O, X_H, T ; K);
 \end{equation}
 \item[M-step:] Maximization of \eqref{eq:CondExpCompLik} with respect to $K$ to update $K^h$ into $K^{h+1}$.
\end{description}
We now give the details of how those two steps are performed.
\paragraph{E-step.} The conditional expectation of the complete log-likelihood writes 
\begin{eqnarray*}
  & & \Esp_{T | X_O; K^h} \left(\Esp_{X_H | X_O, T} \log p(X_O, X_H, T; K) \right) \\
  & = & \Esp_{T | X_O; K^h} \left(\log p(T) + \Esp_{X_H | X_O, T; K^h} \left[\log p(X_O, X_H| T; K) \right] \right).
\end{eqnarray*}
Thanks to the tree structure of the graphical model, we have a simple form for the latter term:
\begin{eqnarray*}
  \Esp_{X_H | X_O, T; K^h} \left[\log p(X_O, X_H| T; K) \right] = \sum_{\{i, j\} \in T} p_{ij}(K),
\end{eqnarray*}
where 
$p_{ij}(K)$ is $-2K_{ij}\widehat{\Sigma}_{ij}$ if both $i \neq j$ are observed, $2K_{ij} W_{ij}^h$ if $i$ is observed and $j$ is hidden, $-K_{ii} \widehat{\Sigma}_{ii}$ if $i = j$ is observed and $-K_{ii} B_{ii}^h$ if $i = j$ is hidden, 
variance and covariance matrices being given by
\begin{eqnarray*}
  W_{HO}^h & = & (K_H^h)^{-1} K_{HO}^h \widehat{\Sigma}_O, \\
  V_{H}^h & = & (K_H^h)^{-1} K_{HO}^h \widehat{\Sigma}_O K_{OH}^h (K_H^h)^{-1}, \\
  B_{H}^h & = & (K_H^h)^{-1} + V_{H}^h.
\end{eqnarray*}
As explained in Section \ref{model}, the diagonal term $K_{ii}$ should actually depend on the tree $T$. We  work here with a common parameter $K_{ii}$, which may result in non-positive definite matrices $K_T$. To circumvent this issue, we project the estimated matrix $K$ on the cone of positive definite matrices at each step of the EM algorithm. In the case where the tree $T$ is supposed to be fixed, the calculation of the conditional distribution \eqref{eq:pTXo} is replaced by the determination of the conditionally most probable tree, likewise in the classification EM introduced by \cite{celeux1992classification}.

\paragraph{M-step.}
Combined with $p(T) \propto \prod_{\{i, j\} \in T} \pi_{ij}$ and with the conditional distribution of $T$, $p(T |X_O; K^h) \propto \prod_{\{i, j\} \in T} \gamma_{ij}$ given in \eqref{eq:pTXo} (with $\gamma_{ij} = \pi_{ij} d_{ij} m_{ij}$), we get that
\begin{eqnarray*}
  \Esp_{X_H, T | X_O; K^h} \log p(X_O, X_H, T ; K) 
  & \propto \; & \Esp_{X_H, T | X_O; K^h} \left[\sum_{\{i, j\} \in T} \log \pi_{ij} + p_{ij}(K) \right] \\
  & \propto \; & \sum_T \left(\prod_{\{k, \ell\} \in T} \gamma_{k\ell}^h\right) \left[\sum_{\{i, j\} \in T} \log \pi_{ij} + p_{ij}(K) \right] 
\end{eqnarray*}
where the normalizing constant does depend on $K^h$ but not on $K$. Hence, at the M-step we need to maximize with respect to $K$
\begin{equation} \label{eq:ObjFunc}
  \sum_T \left(\prod_{\{k, \ell\} \in T} \gamma_{k\ell}^h\right) \left[\sum_{\{i, j\} \in T} p_{ij}(K) \right] = \sum_{i < j} A_{ij} \; p_{ij}(K) 
\end{equation}
where all $A_{ij} = \sum_{T: \{i, j\} \in T} \left(\prod_{\{k, \ell\} \in T} \gamma_{k\ell}^h\right)$ can be computed in $O((p+r)^3)$ using Theorem 3 from \cite{Kir07}. The resulting update formulas of $K$ are given in Appendix~\ref{appendix-2}.

\paragraph{Initialization.} The behavior of the EM-algorithm is known to strongly depend on its starting point. Our initialization strategy is described in Appendix \ref{app:init}.

\subsection{Edge probability and model selection}

In this section, we derive a series of quantities of interest for practical inference.\\
\textbf{Edge probability.} 
In the perspective of network inference, we need to compute the probability for an edge to be part of the tree given the observed data, that is, for edge $\{k, l\}$,
\begin{equation} \label{eq:alpha}
  \alpha_{kl} := P(\{k, l\} \in T | X_O). 
\end{equation}
This probability can be computed for all edges at a time in $O((p+r)^3)$ thanks to Theorem 3 from \cite{Kir07}. It depends on the marginal distribution of the tree $P(T)$ given in \eqref{eq:pT} parametrized with $\pi_{ij}$, which controls the marginal probability of the edge $p_{ij}^0 := P(\{i, j\} \in E_T)$ in a complex manner. In a decision making perspective, it may be desirable to set this probability to an uninformative value such as $1/2$. This probability change can be achieved in $O(p+r)^2$ \citep{BT}.\\
\textbf{Conditional entropy of the tree.} 
We are also interested in the variability of the distribution of the tree given the observed data, measured by its entropy. Denoting $Z_O$ the normalizing constant of the conditional distribution $P(T|X_O)$, we have that
\begin{eqnarray} \label{eq:EntTsXO}
 H(T|X_O) & = & - \sum_T P(T|X_O) \log P(T|X_O) \nonumber \\
 & = & - \sum_T P(T|X_O)  \left(-\log Z_O + \sum_{kl \in T} \log \gamma_{kl} \right) \nonumber \\ 
 & = & \log Z_O - \sum_{kl} \log \gamma_{kl} \left(\sum_{T: kl \in T} P(T|X_O) \right) \nonumber \\
 & = & \log Z_O - \sum_{kl} \alpha_{kl} \log \gamma_{kl} 
\end{eqnarray}
which can be computed with complexity $O((p+r)^2)$, once the edge probabilities $\alpha_{kl}$ have been computed. \\
Because our model involves two hidden variables ($T$ and $X_H$), one may be interested in the conditional entropy of all hidden variables, that is
\begin{equation*}
H(T, X_H|X_O) = H(T|X_O) + \Esp_{T|X_O} \left[ H(X_H|T, X_O)\right].
\end{equation*}
For the second term, we observe that the conditional distribution of $X_H$ given both $T$ and $X_O$ is a Gaussian distribution with variance $K^{-1}_H$ (which is diagonal), whatever $T$ and $X_O$. As a consequence, $H(X_H|T, X_O)$ is constant, so we get that
\begin{equation*}
\Esp_{T|X_O} \left[ H(X_H|T, X_O)\right] = \frac{r \log(2\pi e)}2 - \frac12 \sum_{i \in H} \log(K_{ii}).
\end{equation*}

\paragraph{Model selection.} 
We now turn to the estimation of the unknown number of hidden nodes $r$. First, a standard Bayesian Information Criterion (BIC) can be defined as $BIC(r) = \log p(X_O; \widehat{K}) - \pen(r)$ where the penalty term depends on the number of independent parameters in $K$, that is
\begin{equation*}
\pen(r) =  \left(\frac{p(p+1)}2 + rp + r\right) \frac{\log n}2.
\end{equation*}
Note that the maximized log-likelihood can be computed as
$$
\log p(X_O; \widehat{K}) = \Esp[\log p(X_O, X_H, T) | X_O; \widehat{K}] + H(X_H, T|X_O, \widehat{K}).
$$
In the context of classification, \cite{BCG00} introduced {an} Integrated Complete Likelihood (ICL) criterion where the conditional entropy of the hidden variable is added to the penalty. The {rationale} behind ICL {is} a preference for models with lower uncertainty for the hidden variables. Because we are mostly interested in network inference, it seems desirable to penalize only for the conditional entropy of the tree. This leads to the following criterion
\begin{equation*}
ICL_T(r) = \log p(X_O; \widehat{K}) - H(T|X_O) - \pen(r)
\end{equation*}
where $H(T|X_O)$ is given by \eqref{eq:EntTsXO}.
In {situations} where a reliable prediction of the hidden node $X_H$ is of interest, both entropies can be used in the penalty leading to
\begin{equation*}
ICL_{T, X_H}(r) = \log p(X_O; \widehat{K}) - H(T, X_H|X_O) - \pen(r). 
\end{equation*}

\section{Numerical Experiments}\label{experiments}

\subsection{Experimental setup}\label{subsec:Synthetic}

\begin{figure}[!h]
\centering
\begin{tabular}{cc}
  \includegraphics[width=.4\textwidth, height=.4\textwidth]{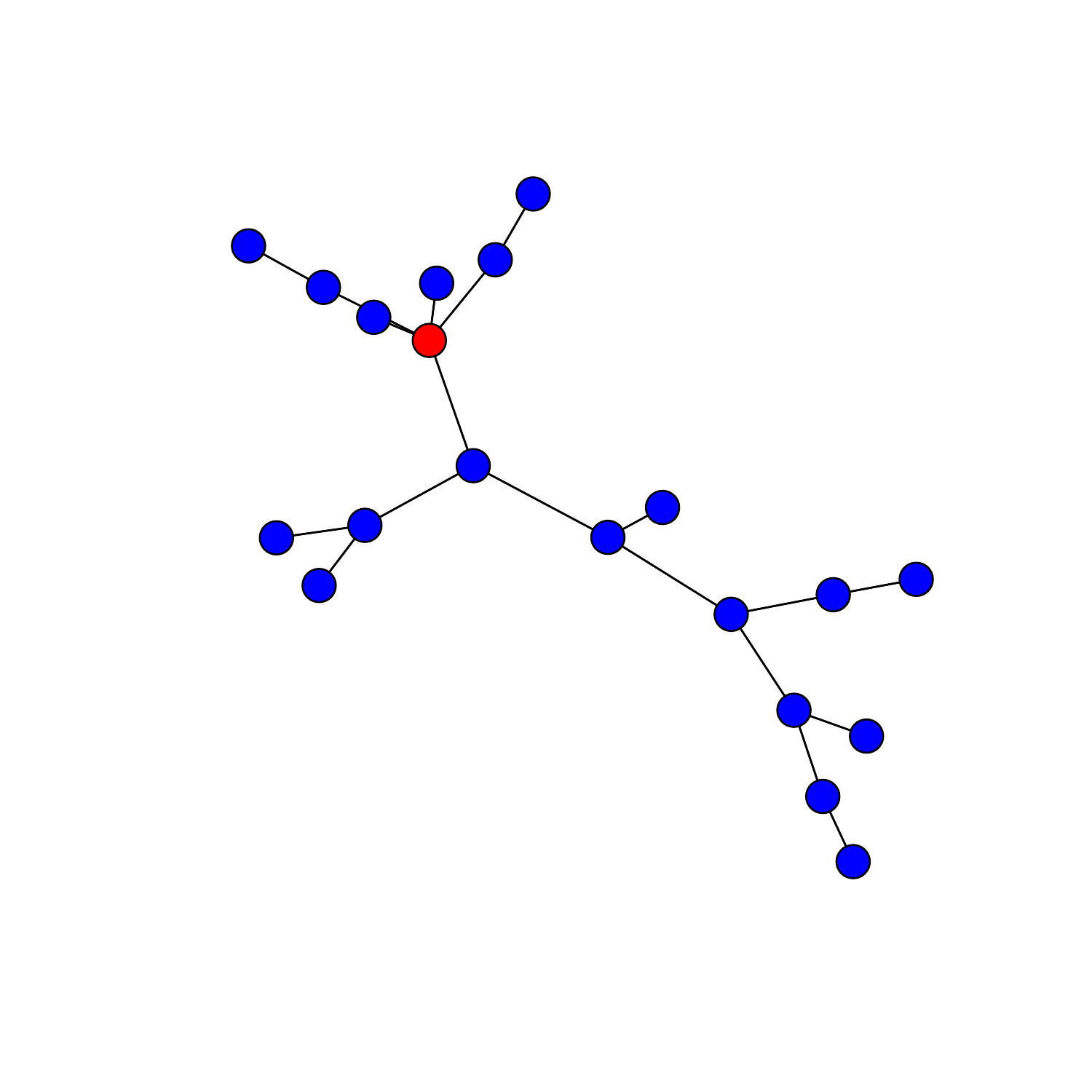} & 
  \includegraphics[width=.4\textwidth, height=.4\textwidth]{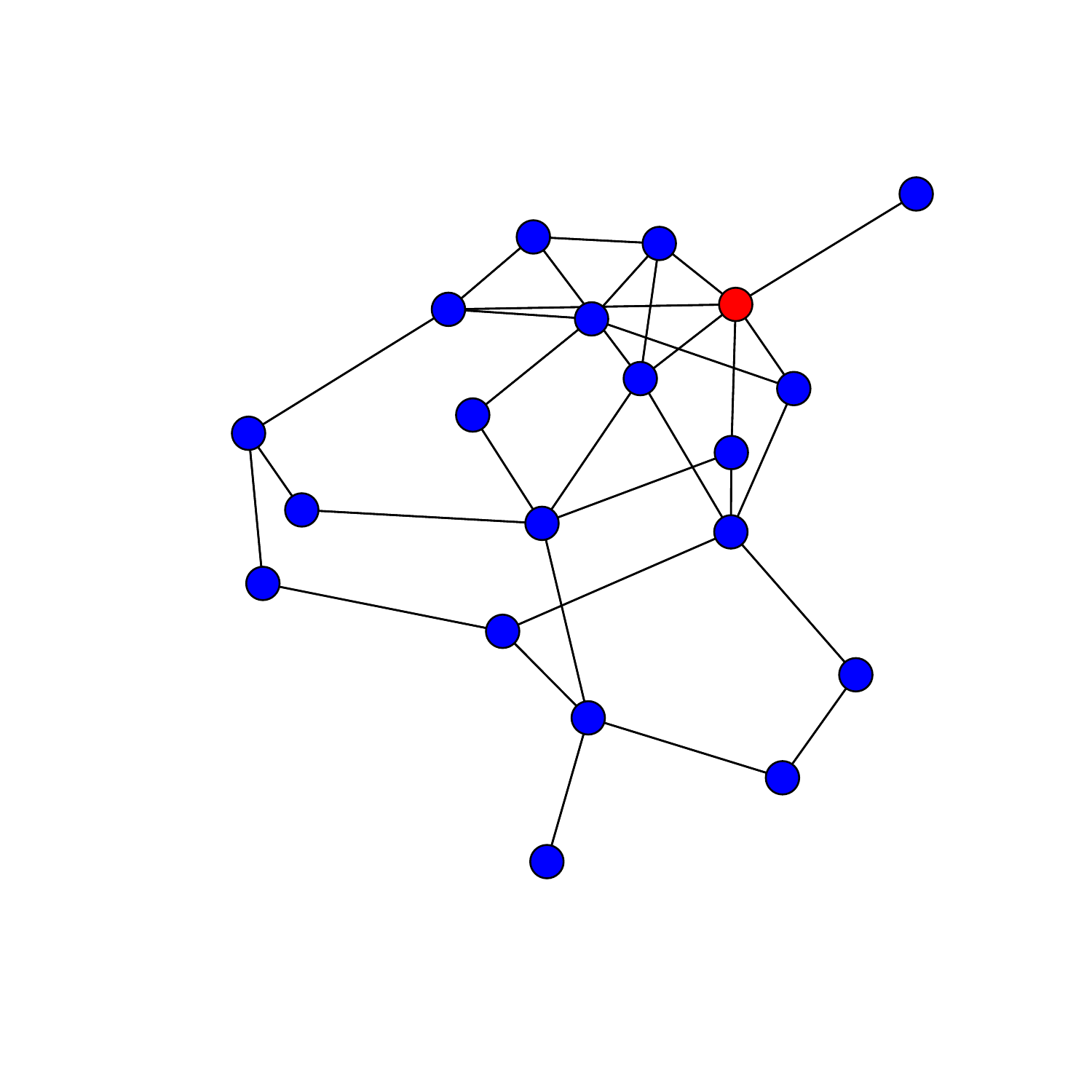}  \\
  (a) Tree and (b) Erd\"os, $p=0.1$ 
\end{tabular}
\caption{Two graph structures used for simulation }
\label{fig:graph-structures}
\end{figure}

Data synthesis in our framework requires the simulation of a graph and of a sparse inverse covariance matrix with matching support. We simulated graphs of two different structures which are given in Figure \ref{fig:graph-structures}, namely a random tree and an Erd\"os-Renyi graph with density 0.1 containing $p=20$ nodes. The binary incidence matrix of  the graph is then transformed  by randomly flipping  the  sign  of  some  elements  in  order  to  simulate  both positively and negatively correlated variables.  Positive definiteness of this  precision matrix $K$ is ensured  by adding a  large enough constant  to the diagonal. We choose the missing nodes at random among those that satisfy the identifiability conditions described in Section \ref{model}.
 The difficulty of detecting missing edges is related to the value of the correlations between the missing nodes and their children. Recall that the marginal precision matrix writes
$$K_m = K_O - K_{OH}K_H^{-1}K_{HO}.$$
We measure the difficulty of detecting the second term $K_{OH}K_H^{-1}K_{HO}$ with the ratio 
$$
SNR = \frac{\left\| K_{OH}K_H^{-1}K_{HO}\right\|_{2}^2}{\left\|K_O\right\|_{2}^2}.
$$
As it increases, the amplitude of the signal coming from the marginalized nodes indeed increases compared to the signal coming from the observed nodes. We control this ratio by multiplying terms in the precision matrix by a constant $\varepsilon$ that we vary:
$$K = \begin{pmatrix}
K_O & \varepsilon K_{OH} \\ 
\varepsilon K_{HO} & \varepsilon K_H
\end{pmatrix}. $$
In the experiments we will consider two settings where $\varepsilon \in \{1, 10\}$. A  Gaussian sample  of size  $n=30$ with zero mean and the above concentration matrix is then simulated 50 times; the results we present below are averaged over the 50 samples. The total complexity of our inference method is $O(n(p+r)^3)$, where $r$ is the (fixed) number of missing nodes. To simulate marginalization, we simply remove in all samples the chosen variable.

\subsection{Edge detection}\label{subsec:Synthetic-results}

We focus this experiment on the ability to recover existing edges of the network, that is the nonzero entries of the concentration matrix. This is a binary decision problem where the compared algorithms are considered as classifiers. The decision made by a binary classifier can be summarized using four numbers: True Positives ($TP$), False Positive ($FP$), True Negatives ($TN$) and False Negatives ($FN$).  We have chosen to draw ROC curves - power ($\text{power}={TP/(FN+TP)}$) versus false positive rate ($\text{FPR}={FP/(FP+TN)}$) - to display this information and compare how well the methods perform. The performance of five algorithms were tested on all the simulated  graph structures~: 
the Chow-Liu  algorithm \citep{ChowLiu},
the graphical lasso \citep{GLasso} (Glasso), 
the EM of \cite{EMlvggm} (EM-Glasso),
the EM algorithm searching for a fixed unknown tree using Chow-Liu algorithm (EM-Chow-Liu),
and our EM algorithm  for tree aggregation (Tree Aggregation). 
Note that the Chow-Liu and Glasso algorithms do not consider missing variables whereas all four other approaches do. 
We compare all methods in terms of marginal graph inference and only the four methods considering missing nodes in terms full graph inference.
We put a special emphasis on the inclusion of 'spurious' edges - that is, edges resulting from marginalization - in the inferred marginal graph. Technically, spurious edges are edges from the marginal graph linking neighbors of the missing nodes in the full graph. To this aim, we plot the fraction $IS/S$ of included spurious edges ($IS$) among the total number of spurious edges ($S$) versus the density of the inferred graph: $(FP+TP)/[p(p-1)/2]$. The interpretation of this curve differs from ROC. An ideal method would keep $IS/S$ to 0 until the end, meaning that the corresponding curve should pushed down to the bottom right corner.

\begin{figure}
  \centering
  \begin{tabular}{ccc}
    \includegraphics[width=0.3\textwidth, height=0.3\textwidth]{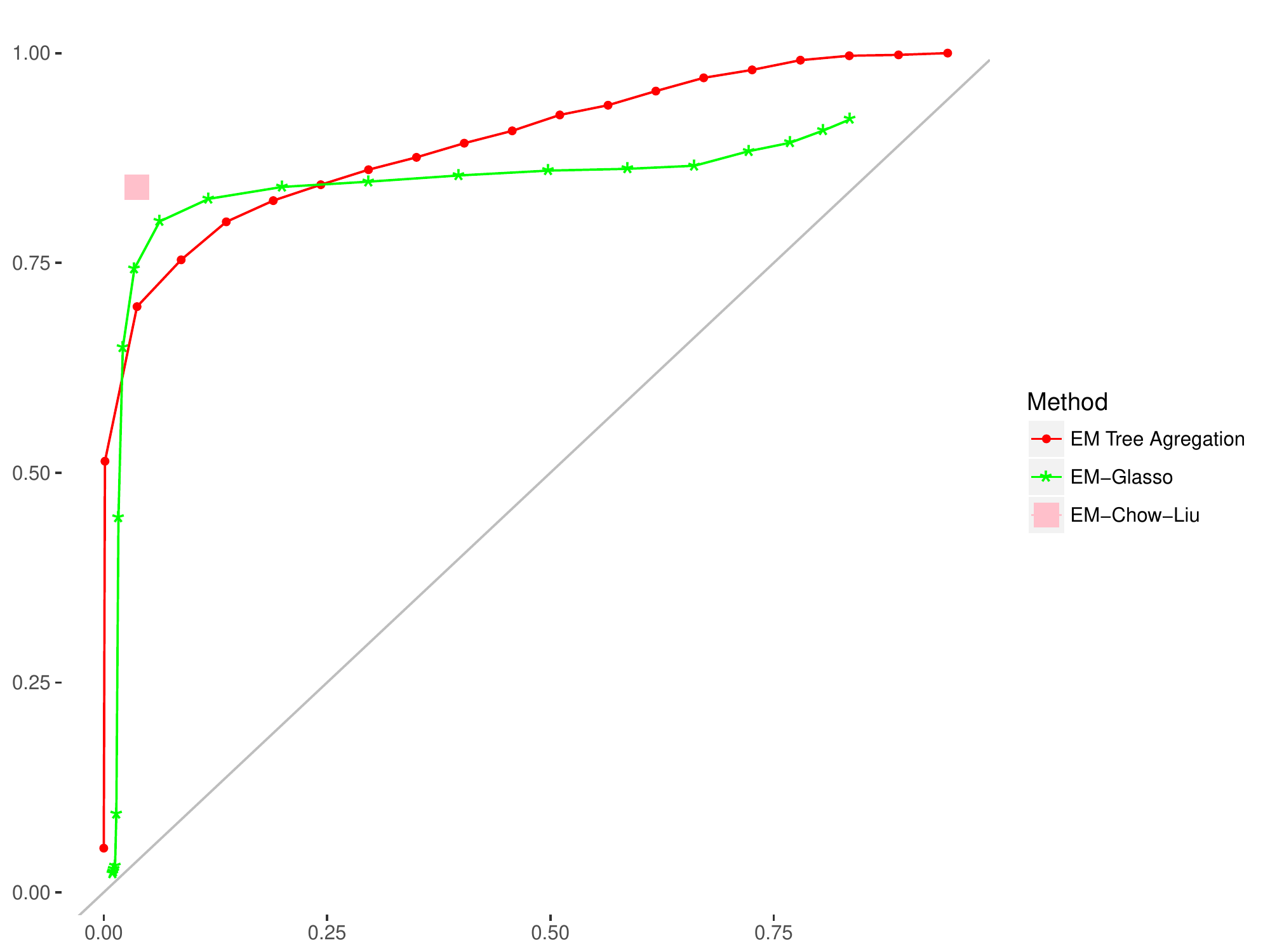} &
    \includegraphics[width=0.3\textwidth, height=0.3\textwidth]{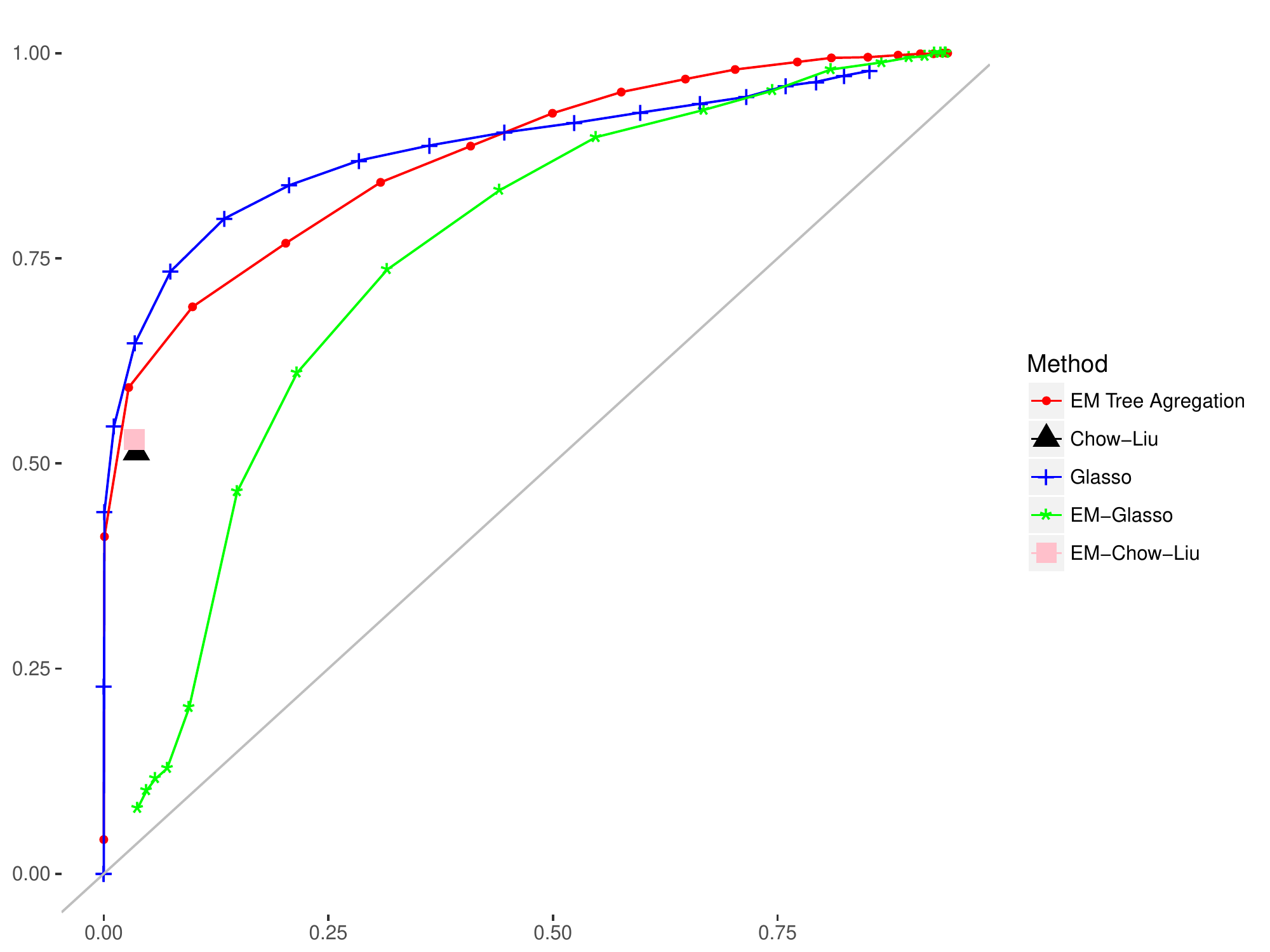} &
    \includegraphics[width=0.3\textwidth, height=0.3\textwidth]{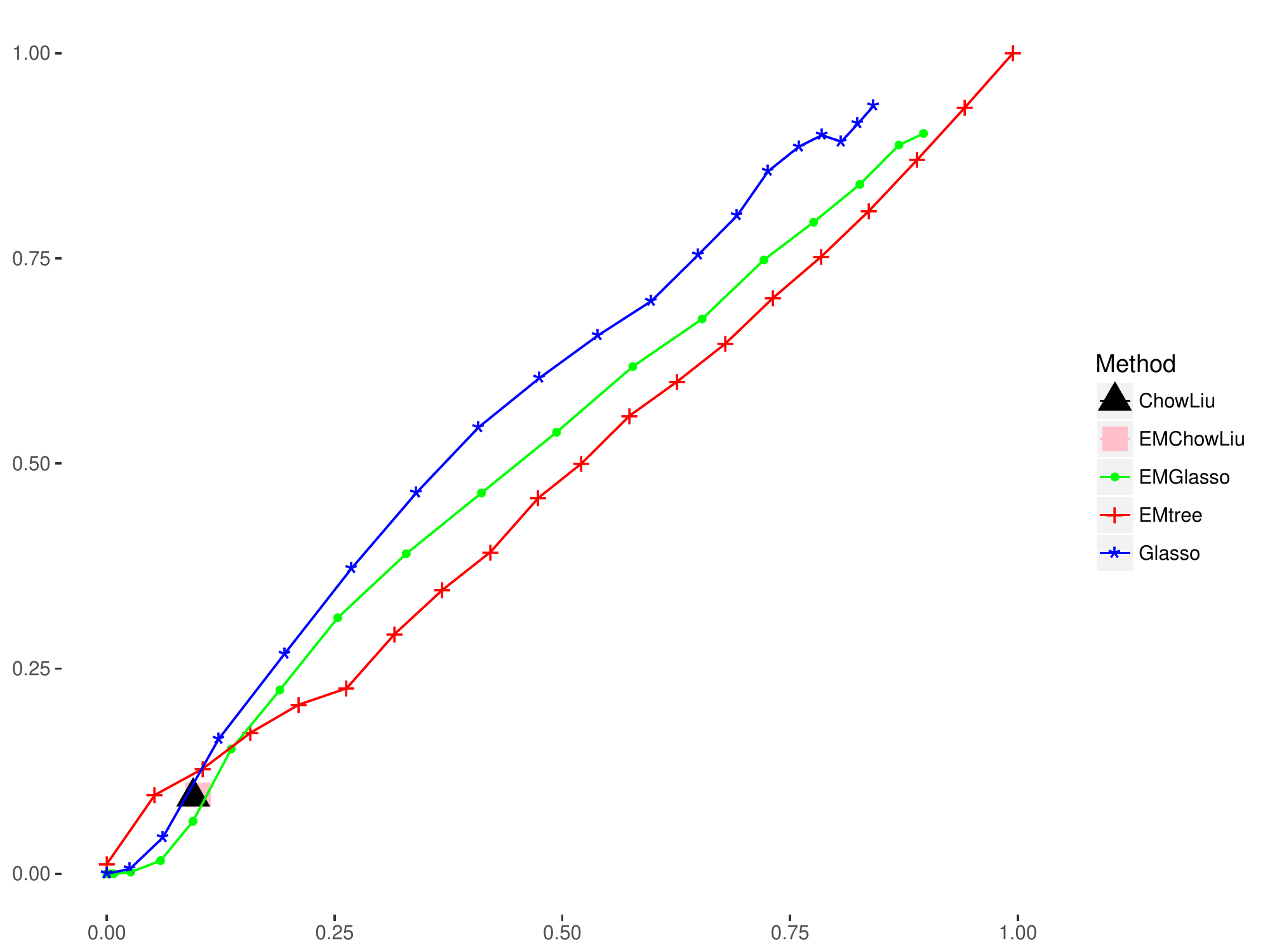} \\
    \includegraphics[width=0.3\textwidth, height=0.3\textwidth]{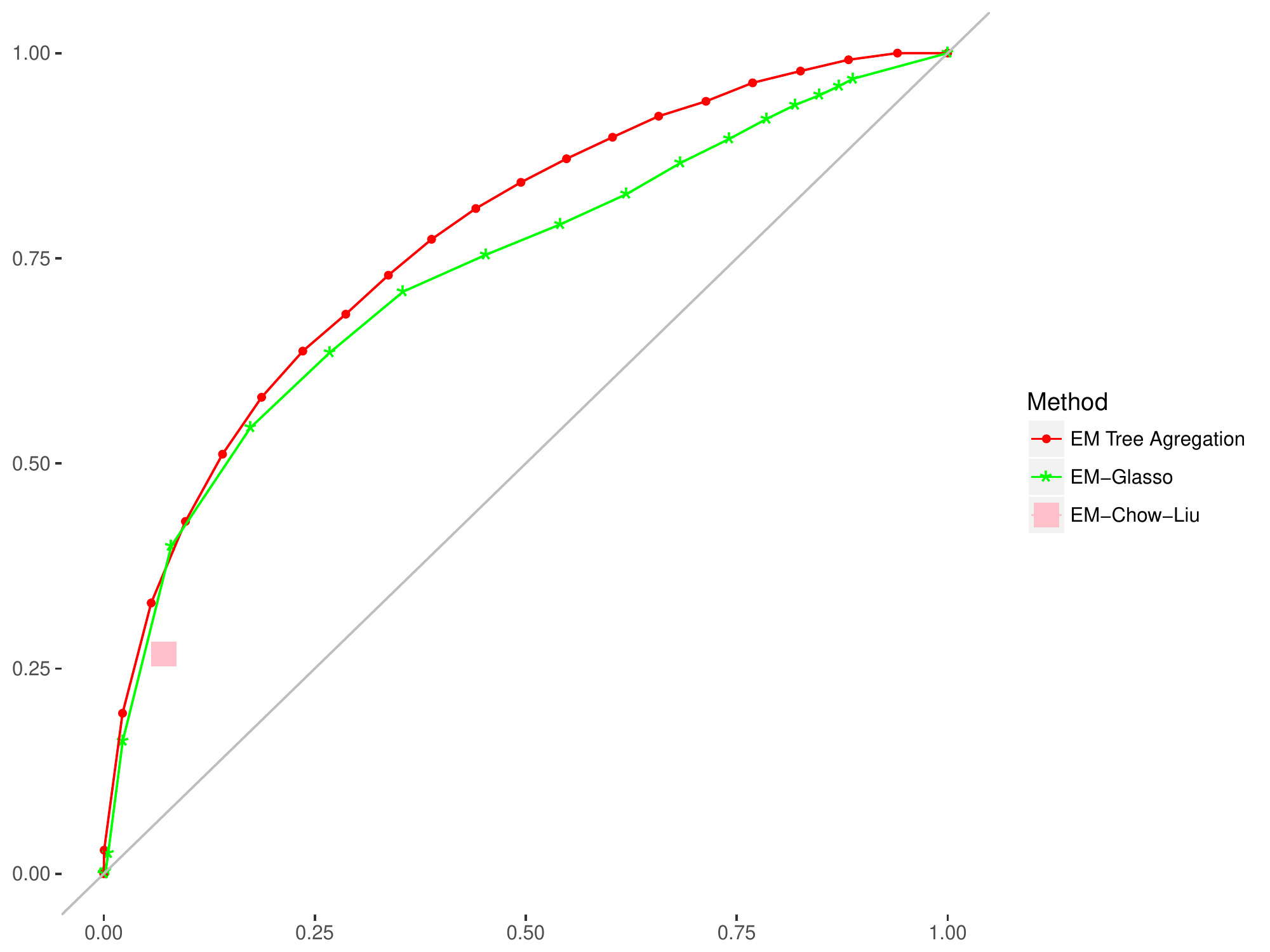} &
    \includegraphics[width=0.3\textwidth, height=0.3\textwidth]{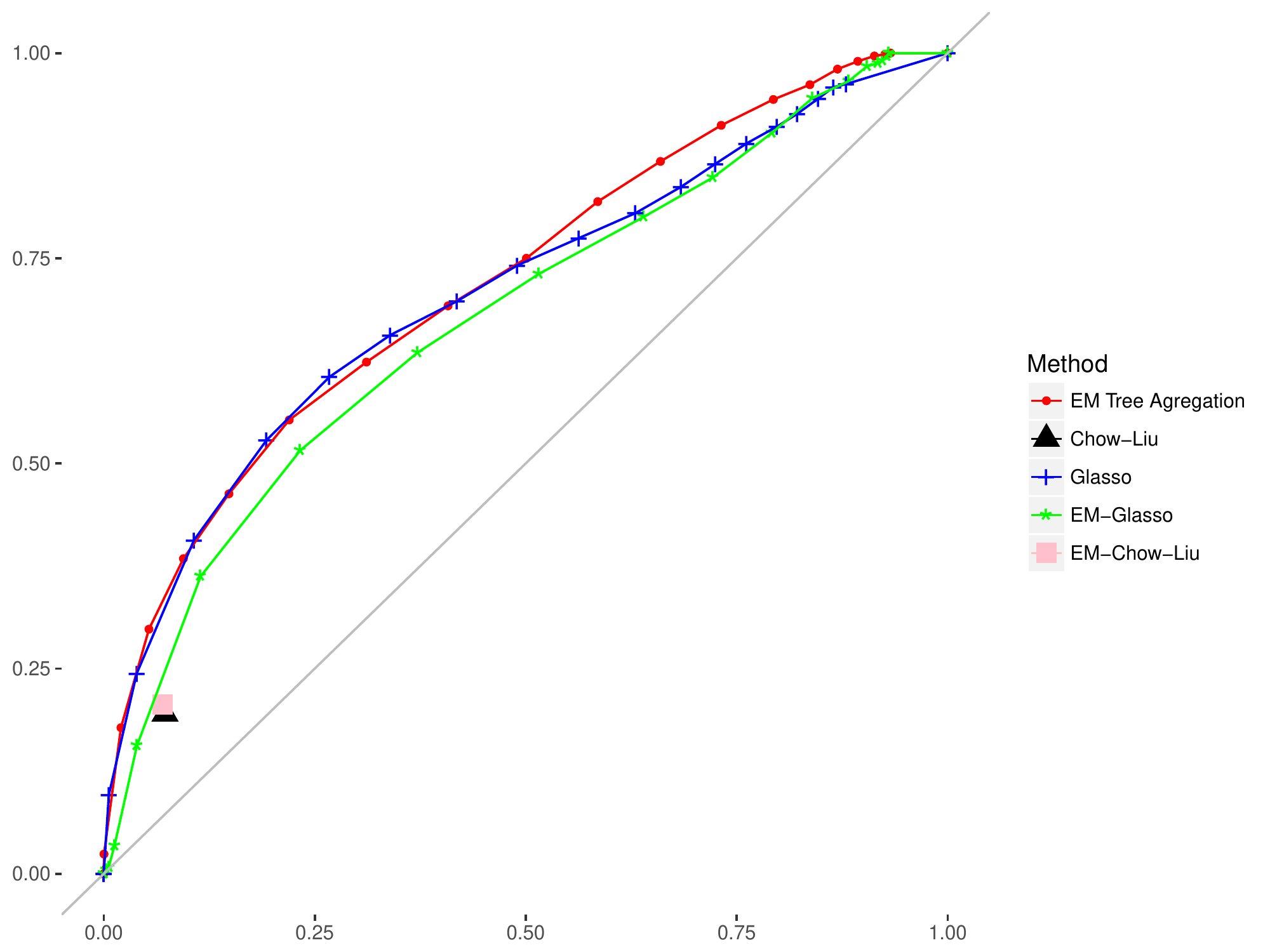} &
    \includegraphics[width=0.3\textwidth, height=0.3\textwidth]{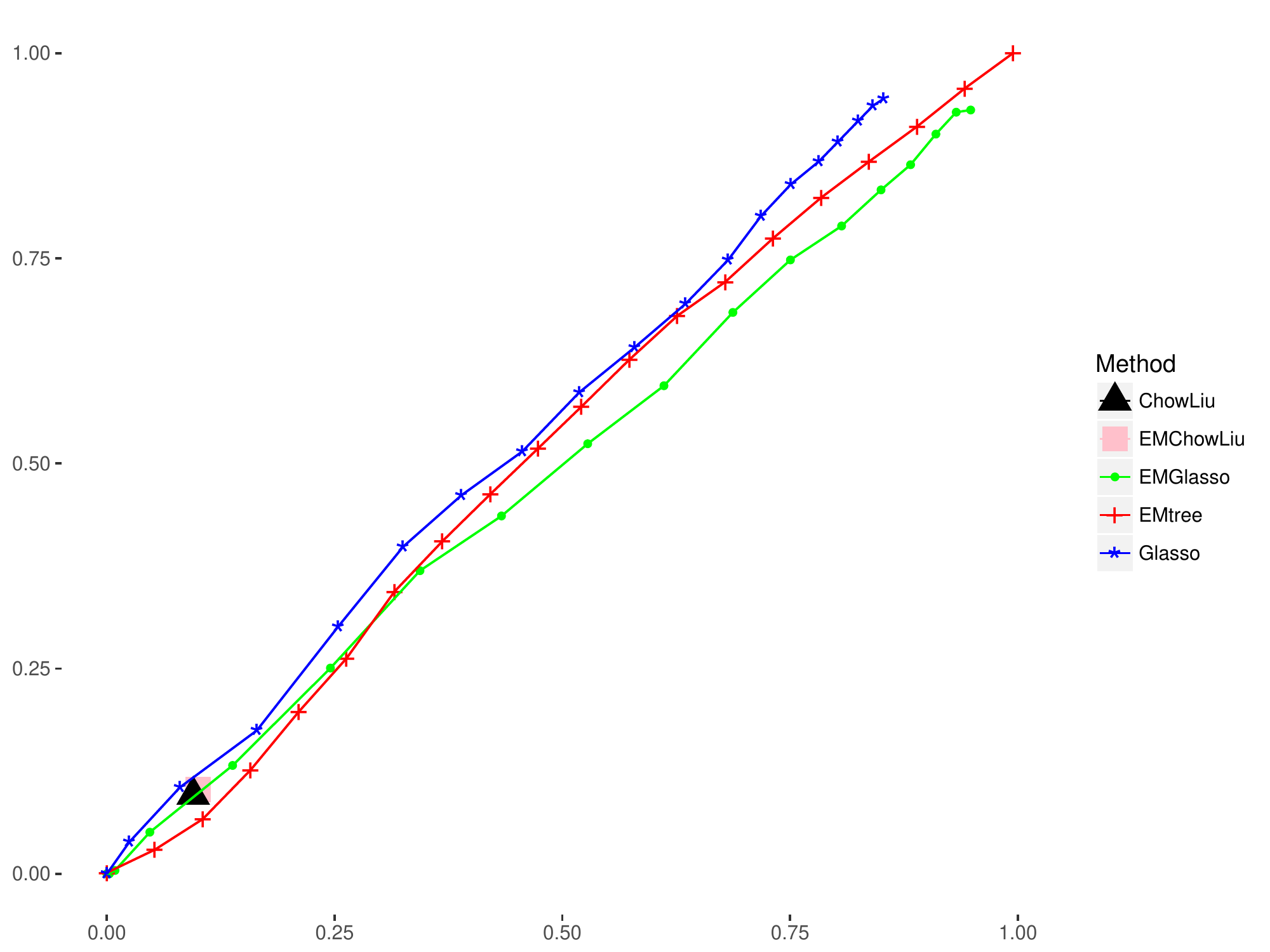} 
  \end{tabular}
  \caption{Simulation results for $SNR=1$. Top: Tree; Bottom: Erd\"os. Left: ROC for the full graph. Center: ROC for the marginal graph; Right: spurious edges.}
  \label{fig:sim-roc-eps1}
\end{figure}

\begin{figure}
  \centering
  \begin{tabular}{ccc}
    \includegraphics[width=0.3\textwidth, height=0.3\textwidth]{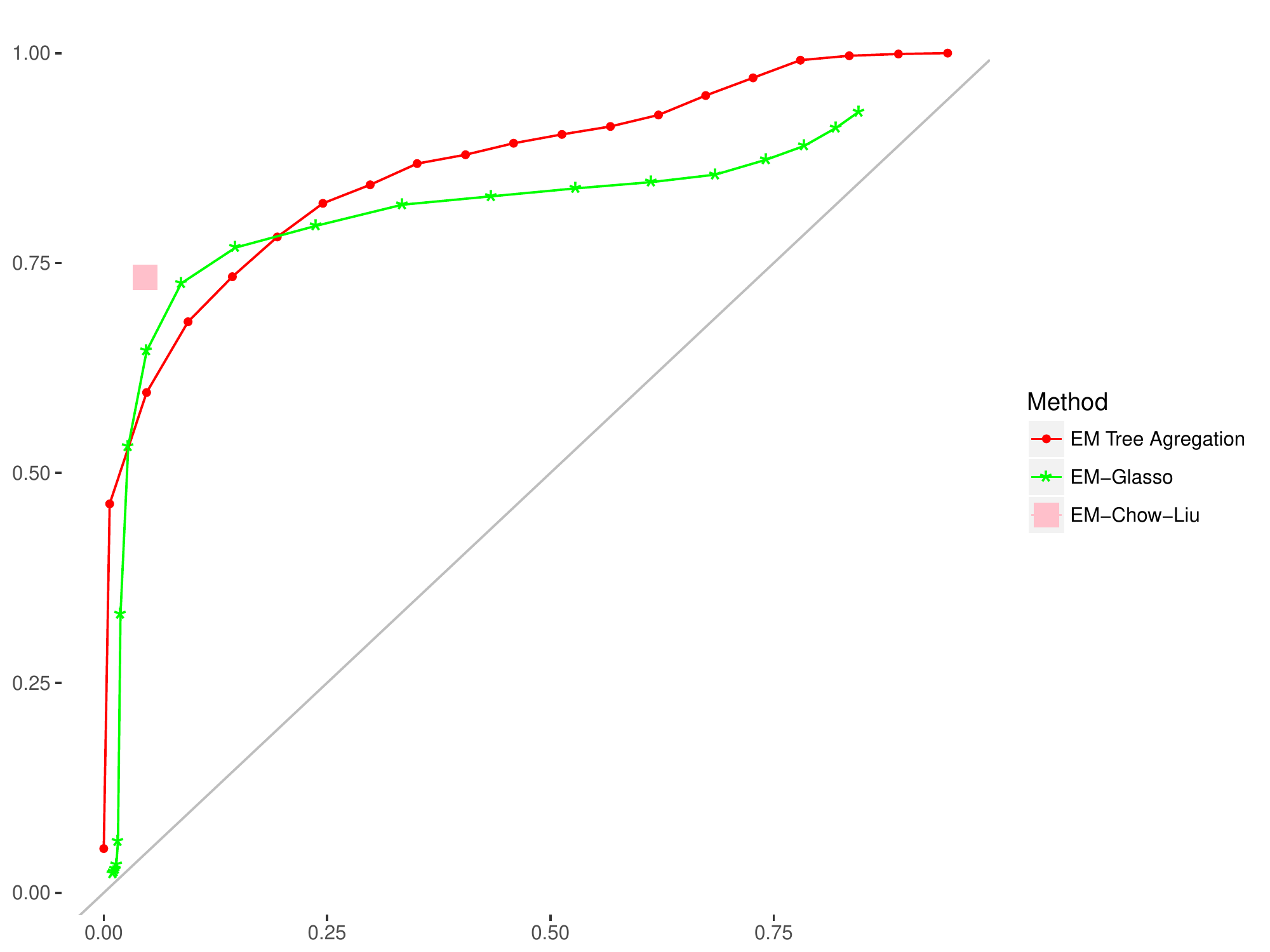} &
    \includegraphics[width=0.3\textwidth, height=0.3\textwidth]{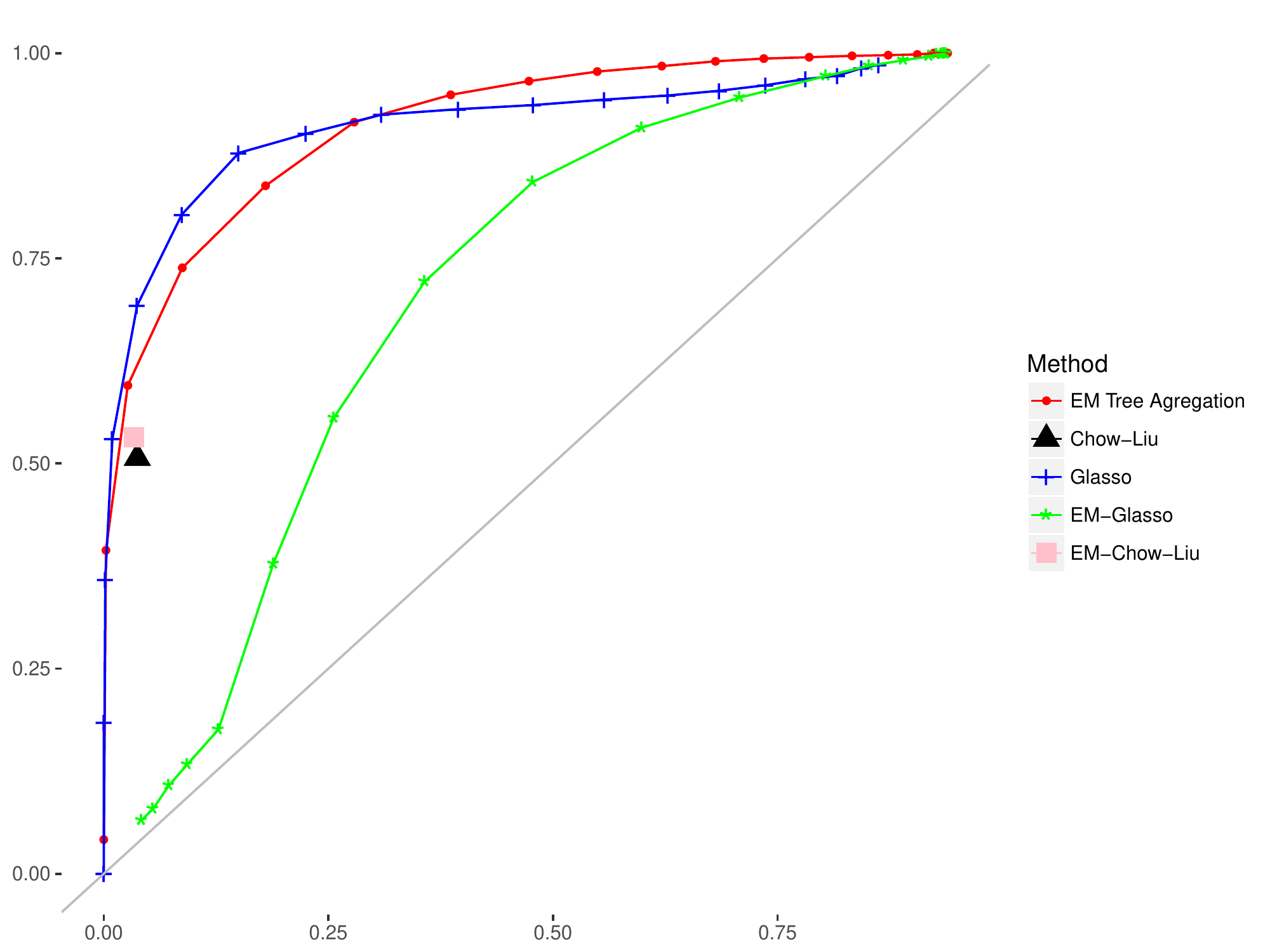} &
    \includegraphics[width=0.3\textwidth, height=0.3\textwidth]{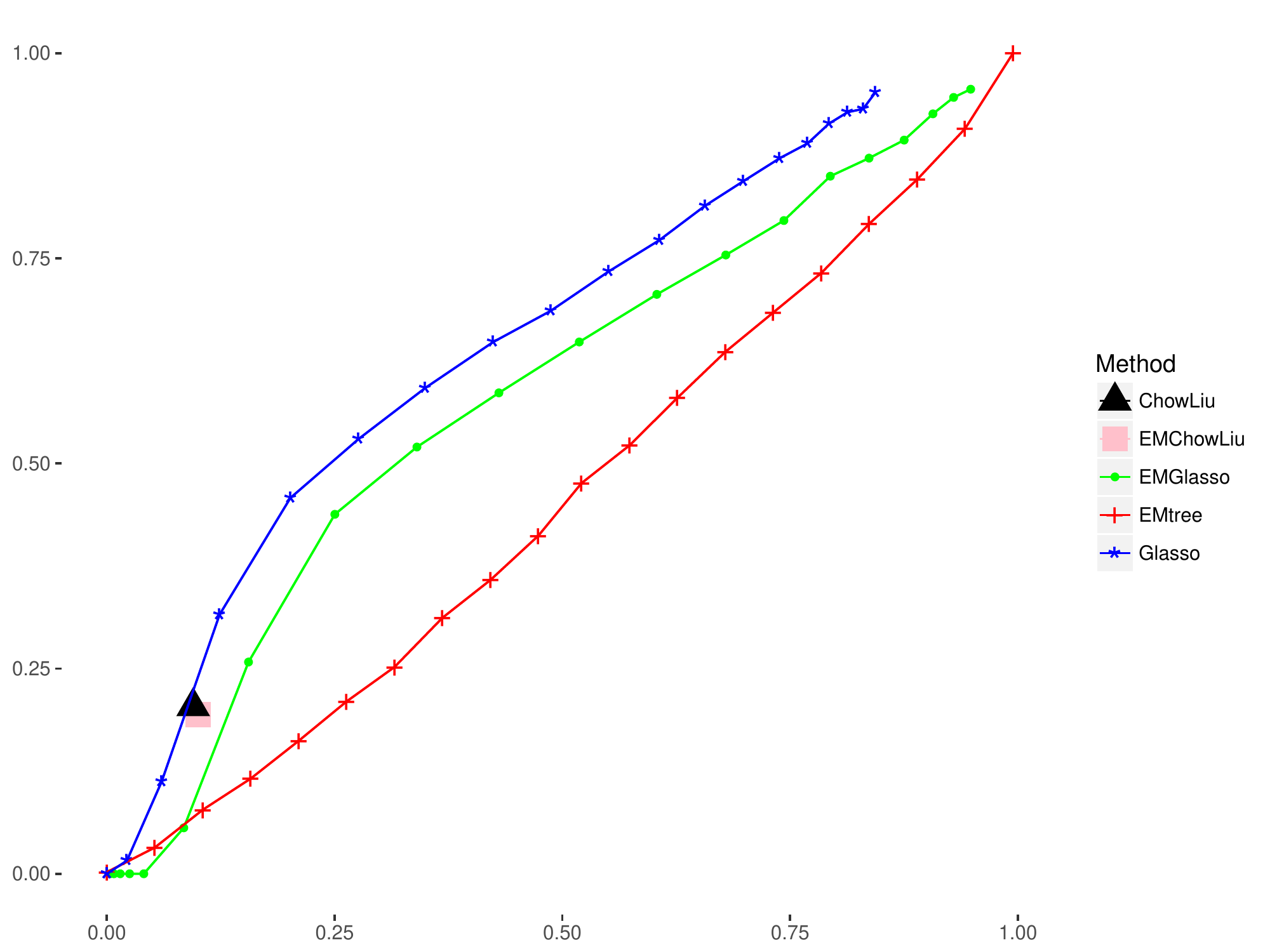} \\
    \includegraphics[width=0.3\textwidth, height=0.3\textwidth]{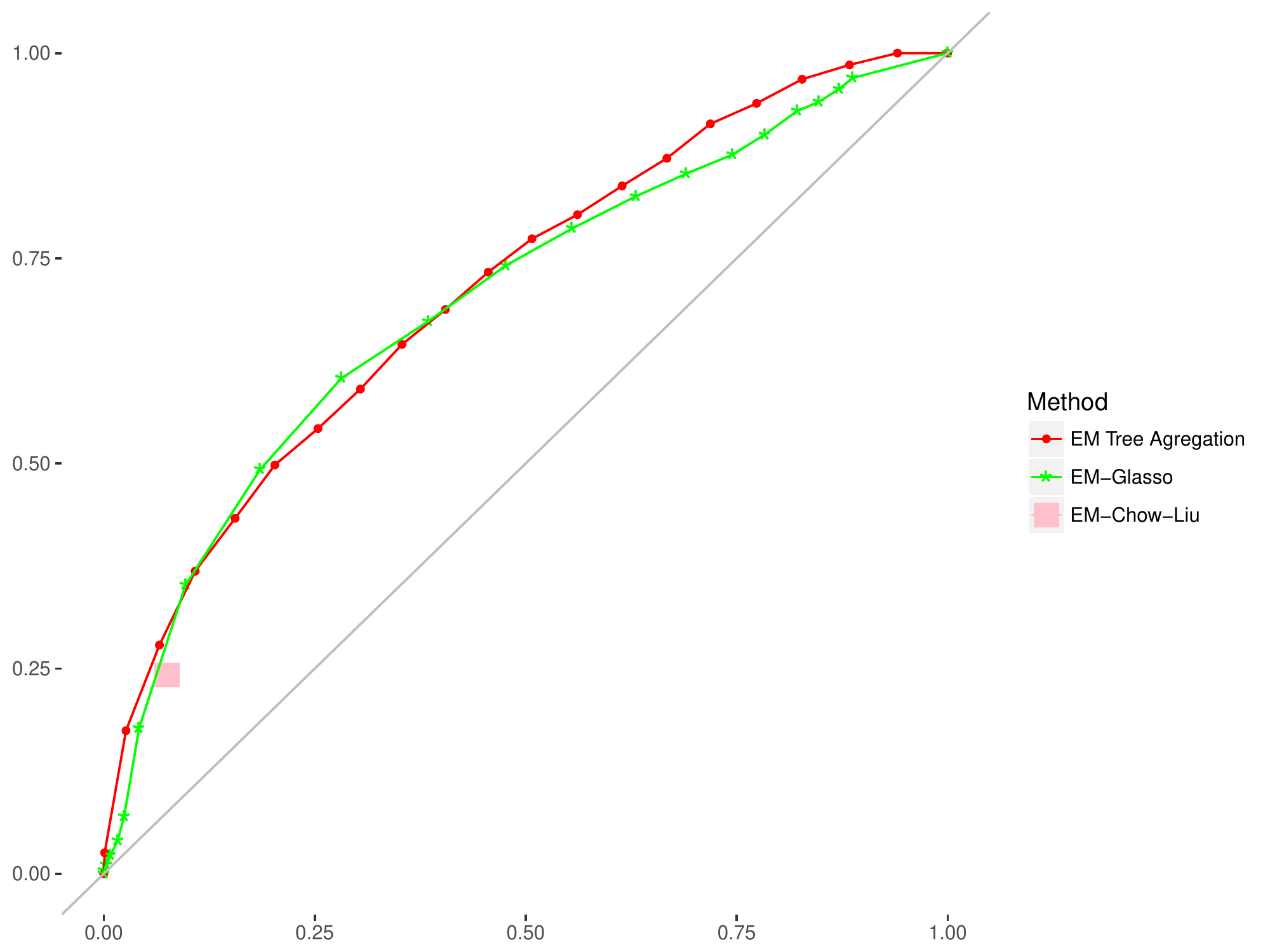} &
    \includegraphics[width=0.3\textwidth, height=0.3\textwidth]{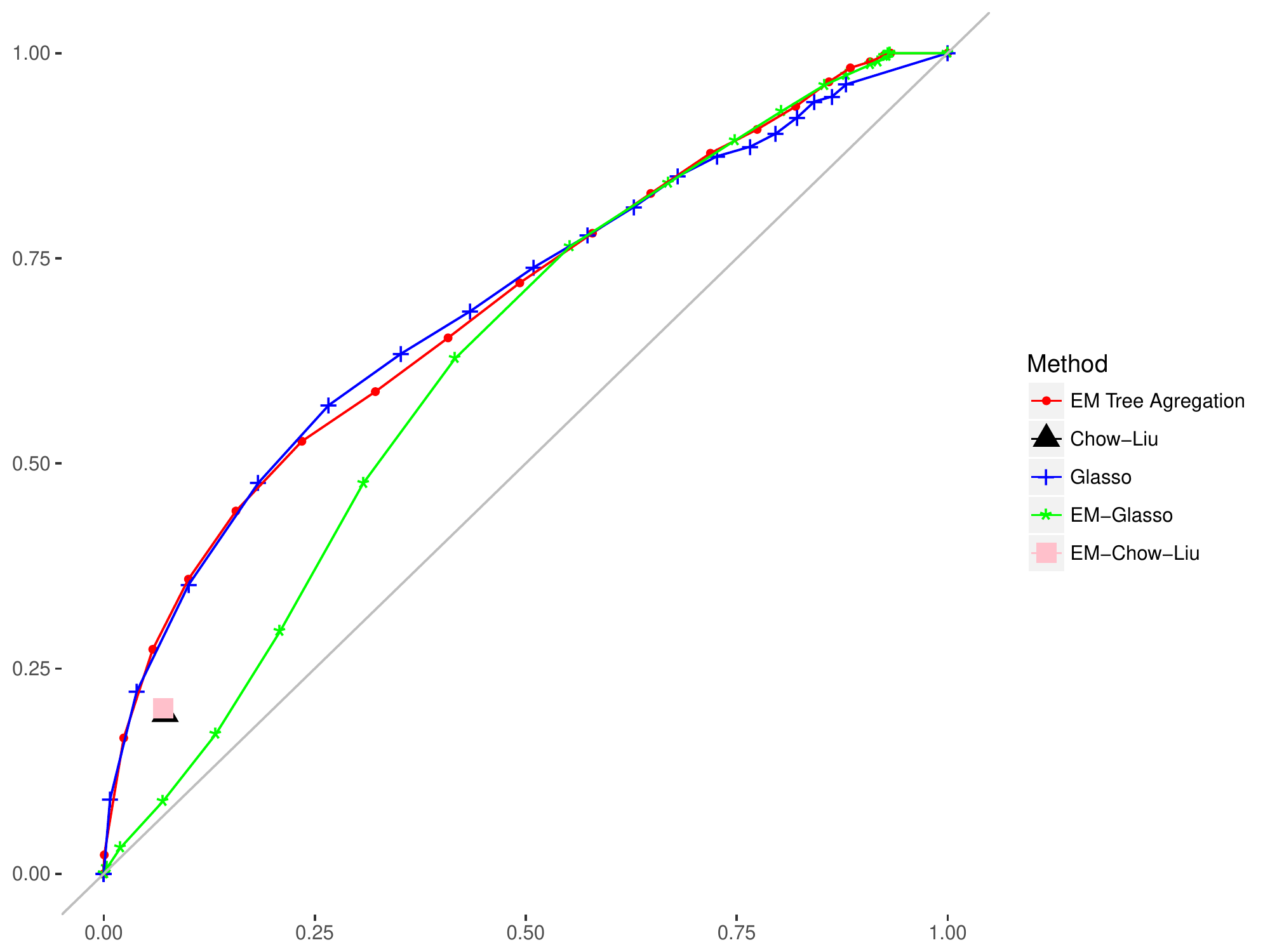} &
    \includegraphics[width=0.3\textwidth, height=0.3\textwidth]{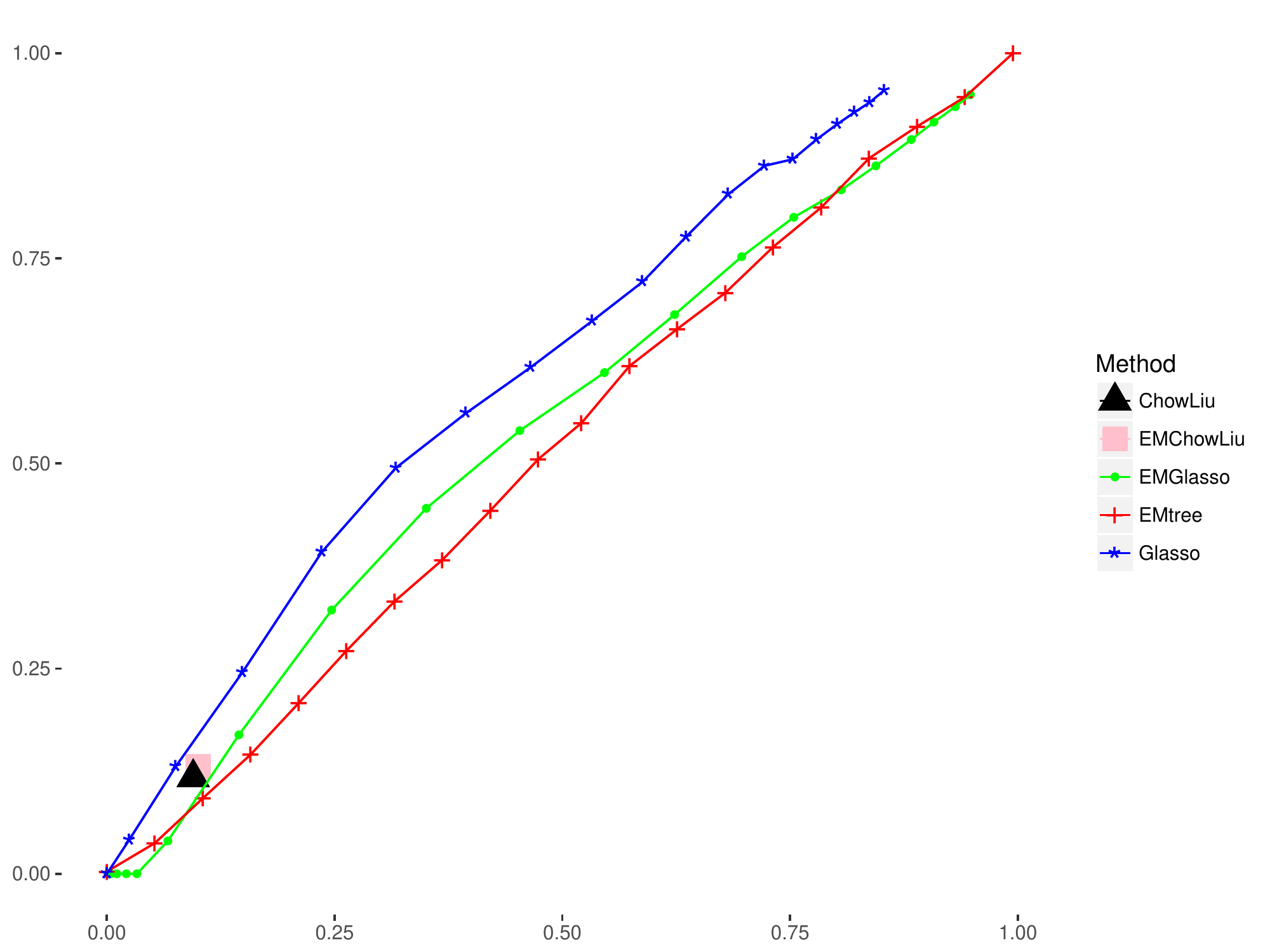} 
  \end{tabular}
  \caption{Simulation results for $SNR=10$. Top: Tree; Bottom: Erd\"os. Left: ROC for the full graph. Center: ROC for the marginal graph; Right: spurious edges.}
  \label{fig:sim-roc-eps10}
\end{figure}
The results are displayed in Figures \ref{fig:sim-roc-eps1} and \ref{fig:sim-roc-eps10}.
The Chow-Liu algorithm and its EM version are very fast to converge and provide very similar solutions of the inference problem. On the marginal graph, even when the true model is a tree, both algorithms do not seem to provide better results than Glasso. 
Glasso and Tree Aggregation perform equally well, and better than EM-Glasso, at inferring the marginal graph.
On the full graph Tree Aggregation performs slightly better than EM-Glasso, which tends to overestimate the number of children of the missing node and therefore has a higher false positive rate. This is in accordance with its underlying model, which assumes that all observed nodes have a hidden parent. Each of these false positive edges in the complete graph induces several false positive edges in the marginal graph. Interestingly, though Tree Aggregation is tailored to infer the full graph, it performs as well as Glasso at predicting the marginal graph, which is the primary target of Glasso.

\subsection{Model selection}\label{subsec:model-selection}
\begin{figure}
  \centering
   \begin{tabular}{c|c}
    \includegraphics[width=.45\textwidth, height=.45\textwidth]{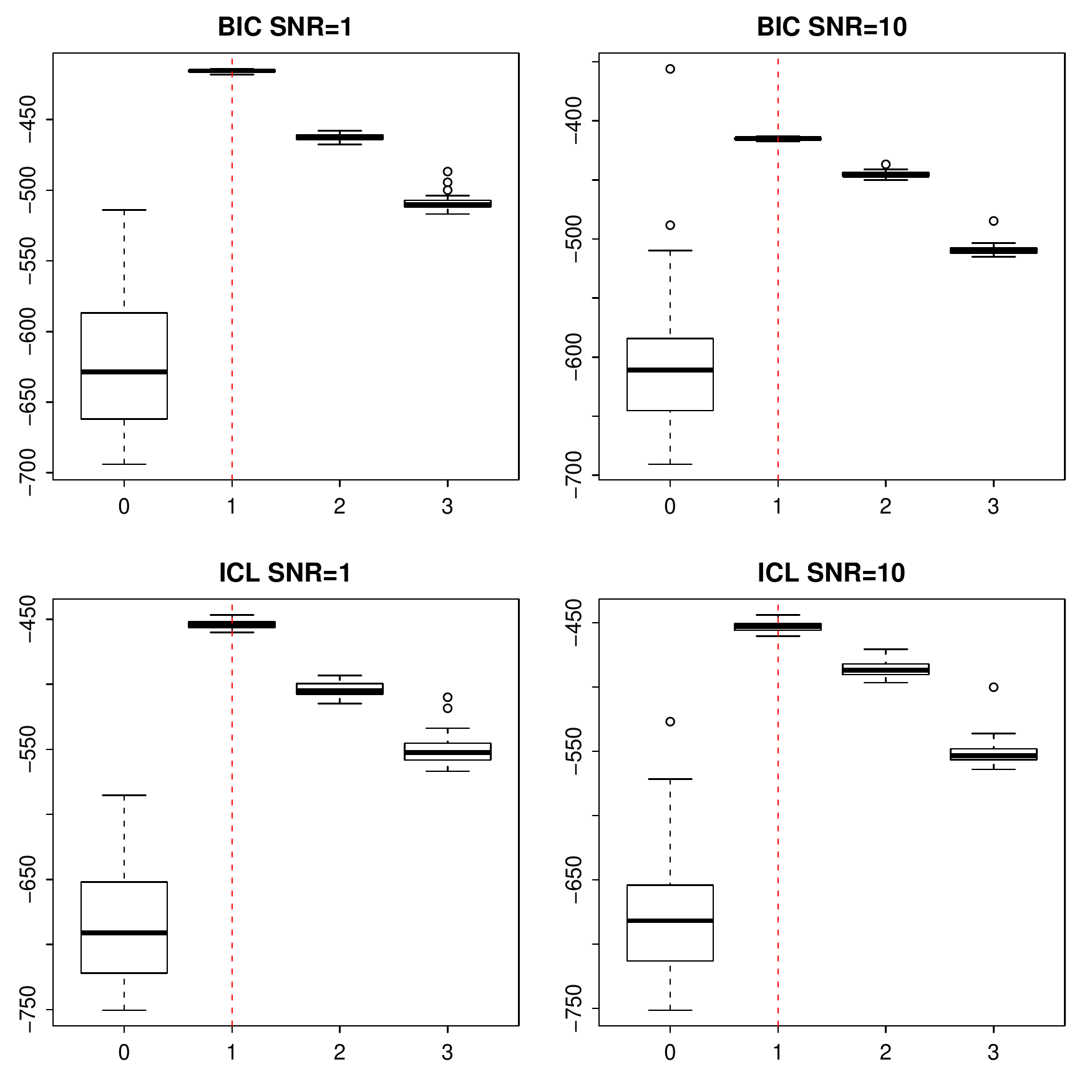} &
    \includegraphics[width=.45\textwidth, height=.45\textwidth]{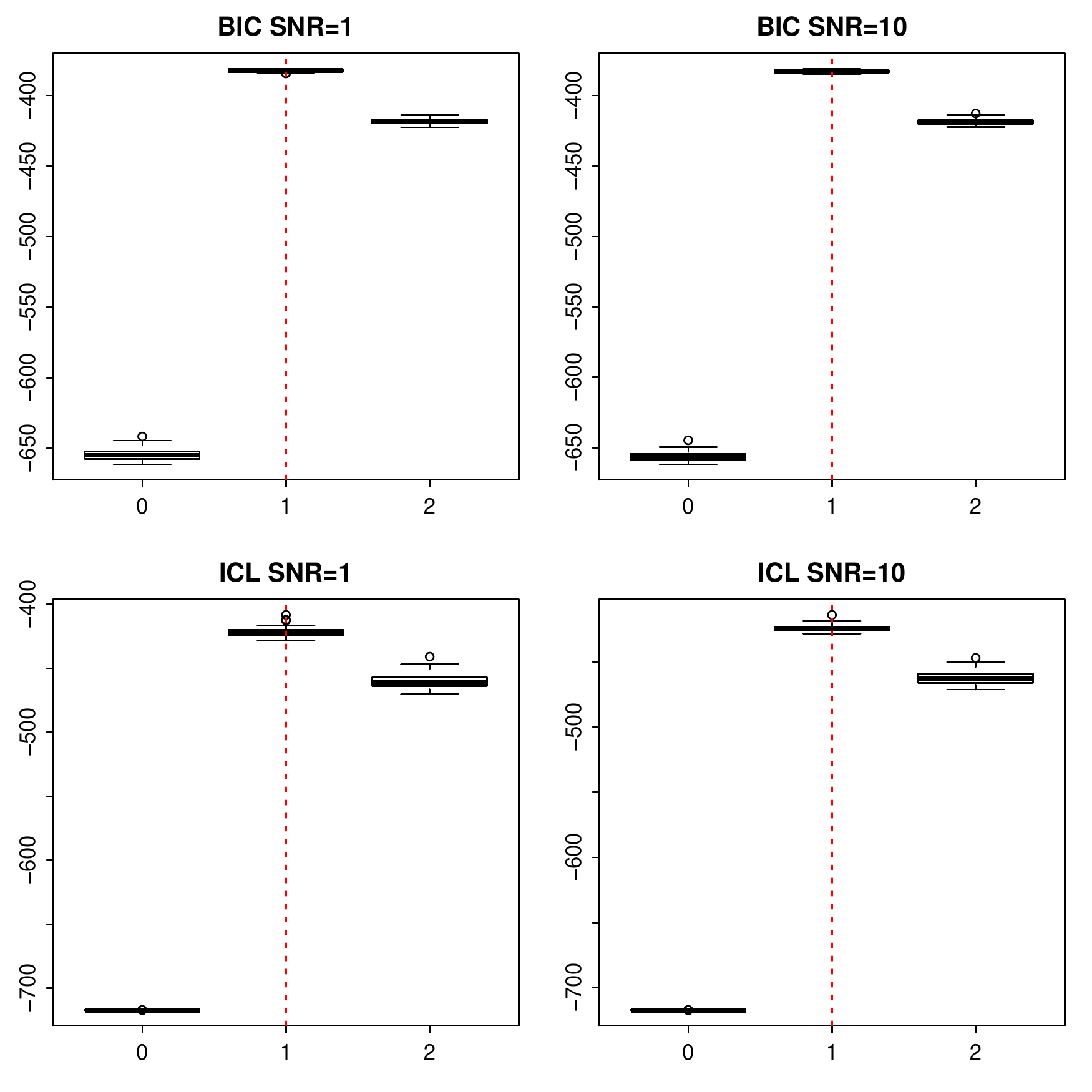} 
   \end{tabular}
  \caption{Model selection. Left block: Tree; Right block: Erd\"os. Top: BIC; Bottom: ICL. Within block left: $SNR=1$, right: $SNR=10$. Dotted red line: true number of missing nodes.}
  \label{fig:sim-bic}
\end{figure}

We now assess the performance of the proposed model selection criteria on the same simulated datasets, in which $r=1$ node is missing. In all simulations, the criteria $ICL_{T, X_H}$ and $ICL_T$ displayed very similar results, the conditional entropy of $X_H$ being very small as compared to this of $T$. As a consequence, we only provide the results for $ICL_T$ (hereafter named simply $ICL$). Figure \ref{fig:sim-bic} shows that, for both network topologies, the BIC and ICL criteria display very similar behaviors and that they all detect the existence of a missing node. When the full network is tree-shaped (Figure \ref{fig:sim-bic}, top), all criteria are maximal for $r=1$, whereas the choice between $r=1$ and $r=2$ is more difficult for the Erd\"os network.

\begin{figure}
  \centering
  \begin{tabular}{c|c}
    \includegraphics[width=.45\textwidth, height=.225\textwidth]{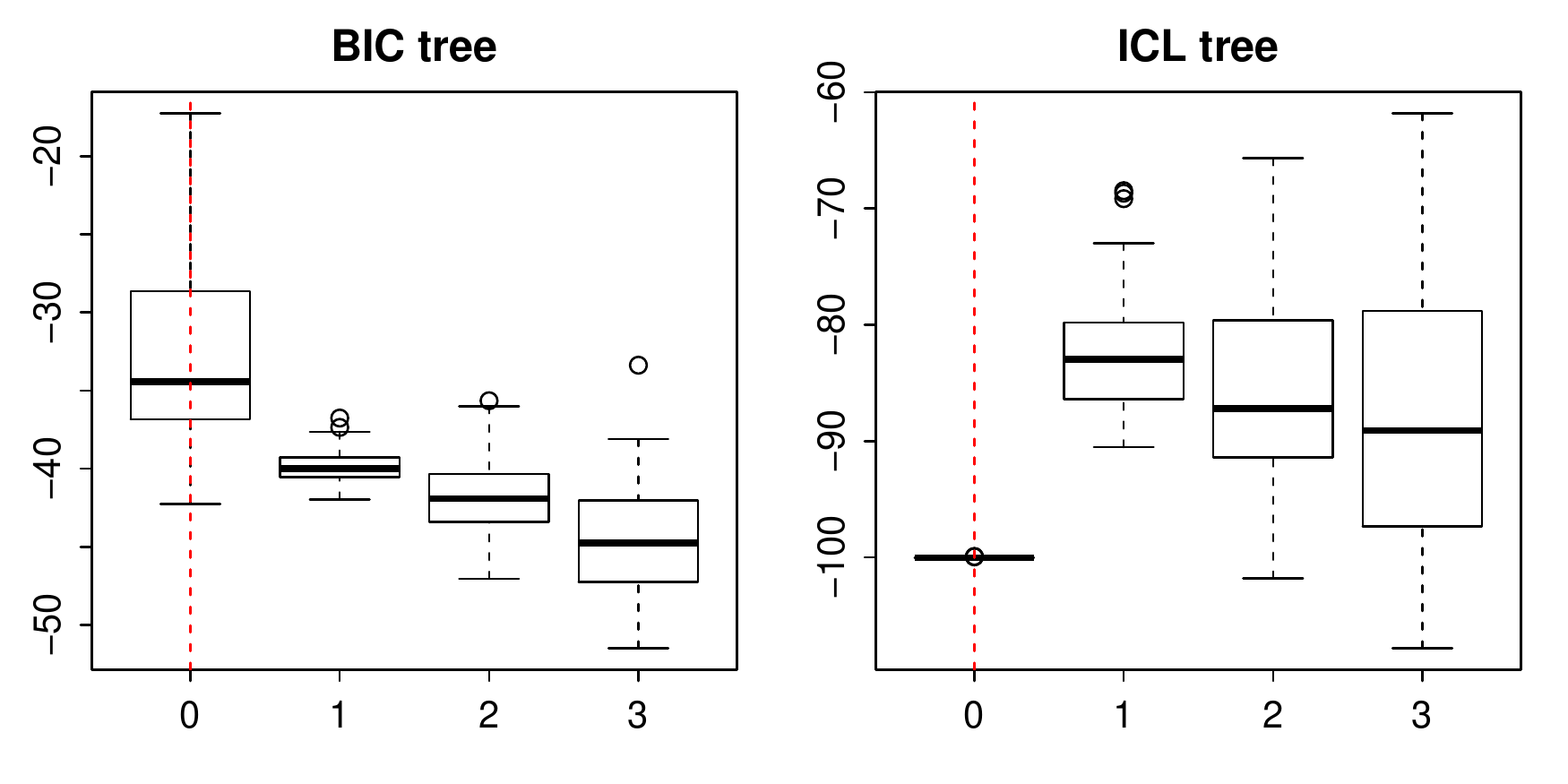} &
    \includegraphics[width=.45\textwidth, height=.225\textwidth]{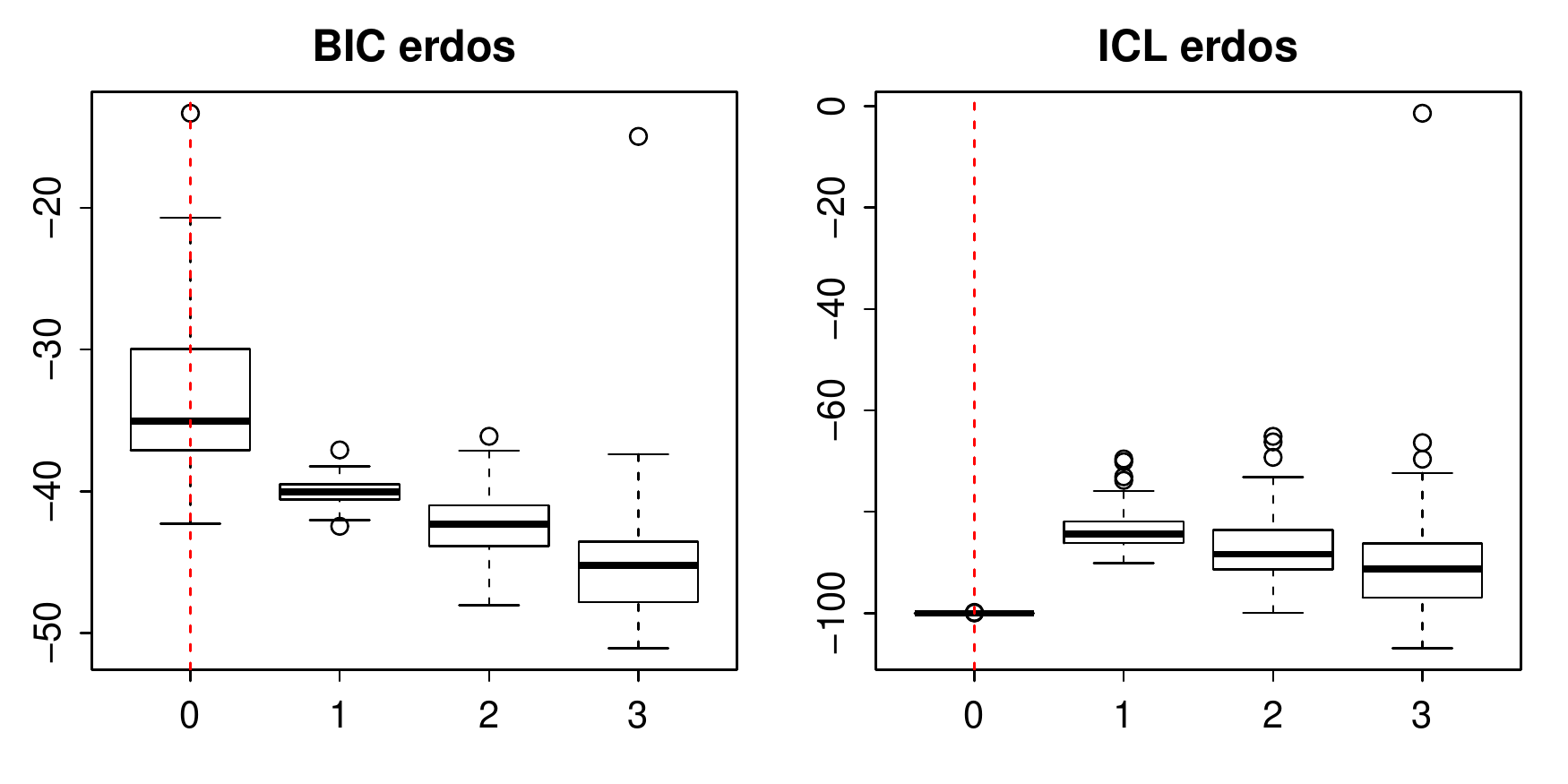} 
   \end{tabular}
  \caption{Model selection. Left block: Tree; Right block: Erd\"os. Within block left: BIC, right: ICL. Dotted red line: true number of missing nodes.}
  \label{fig:sim-bic-pas-cachee}
\end{figure}

We repeat the experiment, this time without marginalizing any node. The results shown in Figure \ref{fig:sim-bic-pas-cachee} show that the BIC criterion doesn't detect any hidden node, contrary to the ICL criterion. Nonetheless the values of ICL for 0, 1, 2 and 3 hidden nodes are much tighter than in the previous example.

\section{Flow cytometry data analysis}\label{subsec:cytoData}
\begin{figure}[!h]
\centering
\begin{tabular}{cc}
  \includegraphics[width=0.4\textwidth, height=0.4\textwidth]{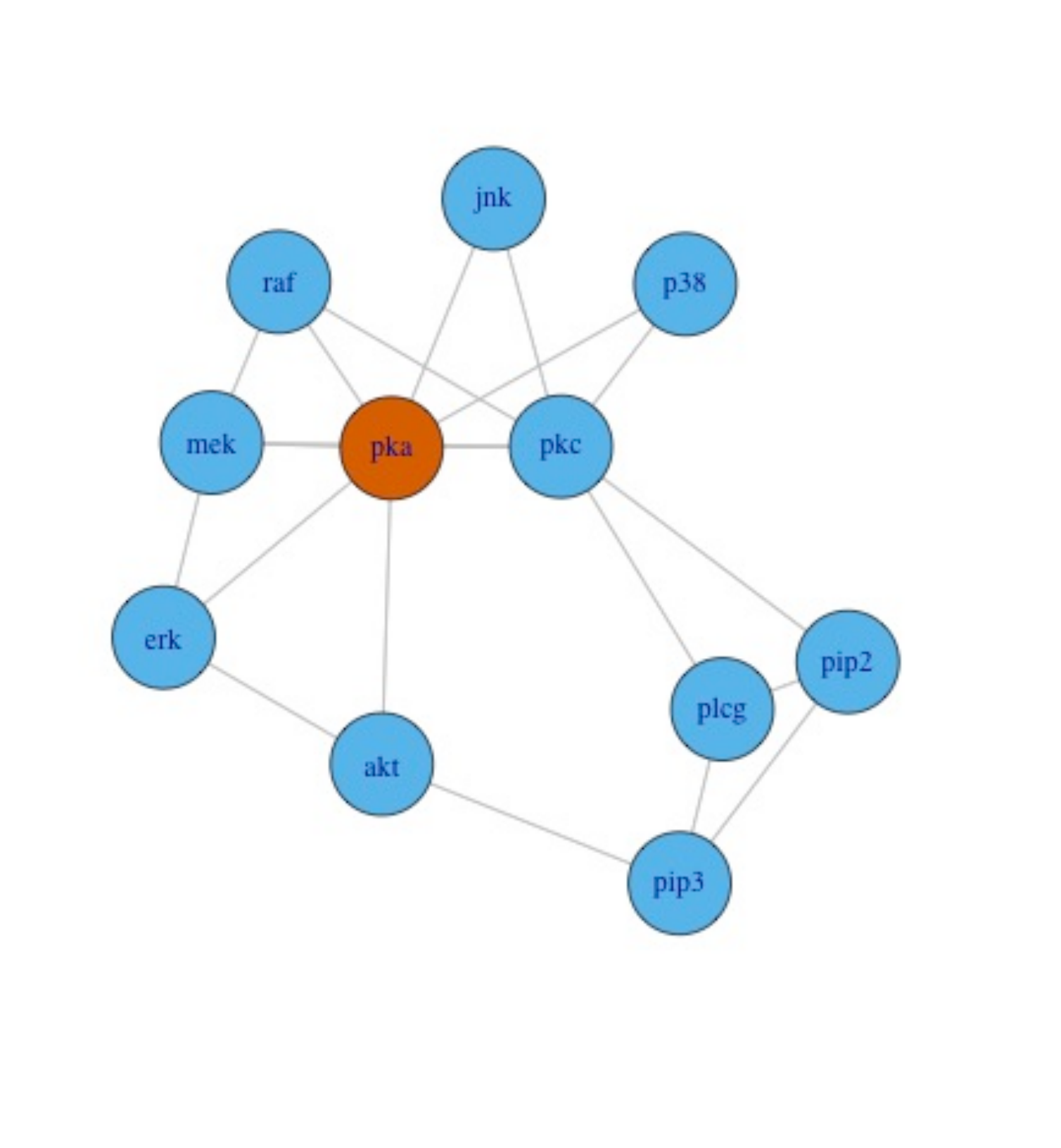}&
  \includegraphics[width=0.4\textwidth, height=0.4\textwidth]{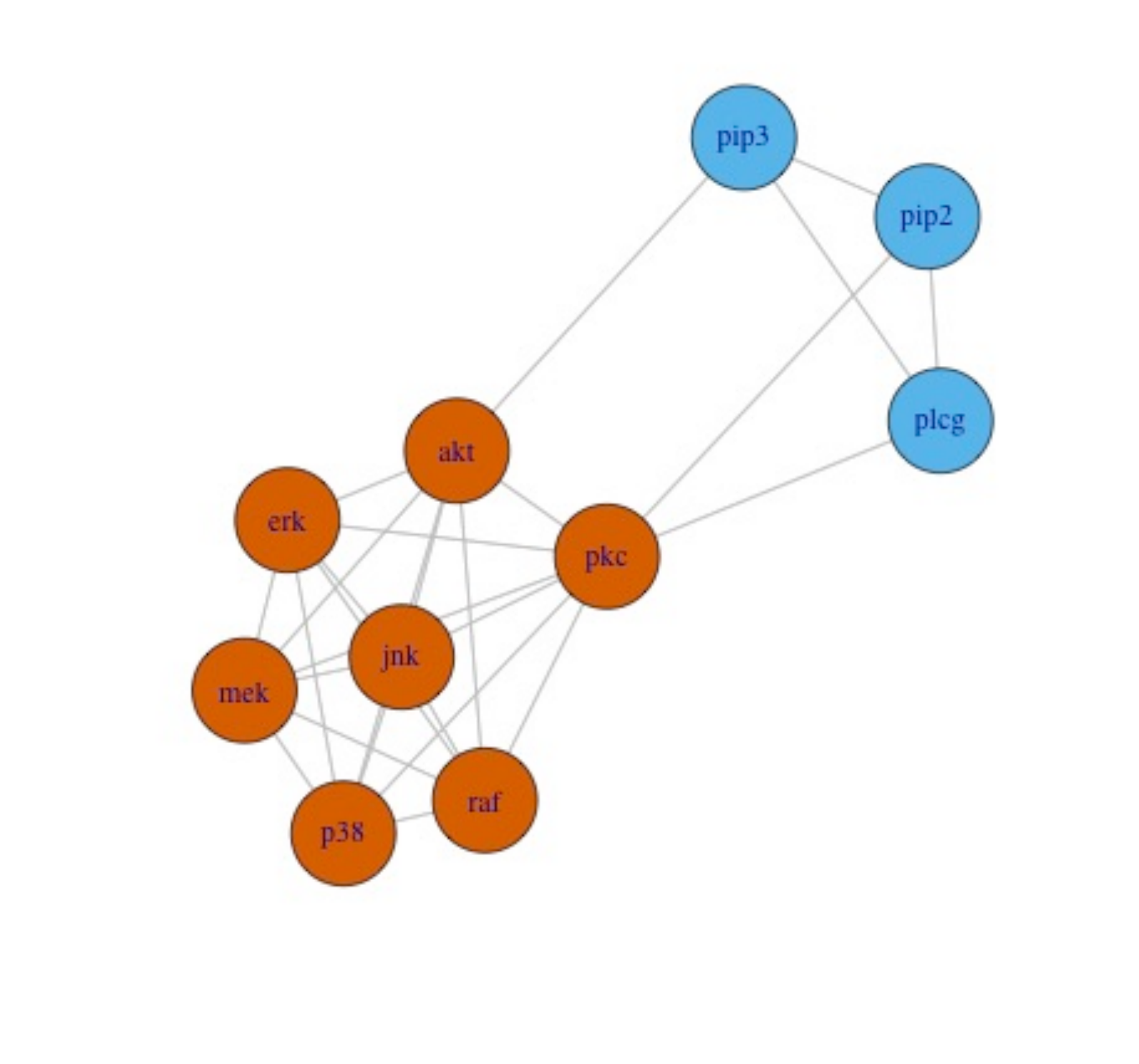}\\
  (a) Full graph (hidden node in red) & (b) Marginal graph 
\end{tabular}
    \caption{Gold standard for Raf pathway}
    \label{fig:raf}
\end{figure}

We applied our procedure to the inference of the Raf cellular signaling network based on flow cytometry data. The Raf network is implied in the regulation of cellular proliferation. The data were collected by \citep{Sachs} and later used by \citep{Werhli} and \citep{BT} in network inference experiments. Flow cytometry measurements consist in sending unique cells suspended in a fluid through a laser beam, and measuring parameters of interest by collecting the light re-emitted by the cell by diffusion or fluorescence. In this study, the parameters of interest are the activation level of $11$ proteins and phospholipids involved in the Raf pathway, and are measured by flow cytometry across $100$ different cells. Though the true structure of this network is unknown, experiments have highlighted a consensus pathway that we used as gold standard to assess the performance of our algorithm. The consensus network displayed in Figure \ref{fig:raf} is far from being a tree. We removed one protein from the dataset, which amounts to hide the corresponding node (in red in Figure \ref{fig:raf}), and applied our algorithm to this marginal data.

\begin{figure}[!h]
\centering
\begin{tabular}{cc}
  \includegraphics[width=0.4\textwidth, height=0.4\textwidth]{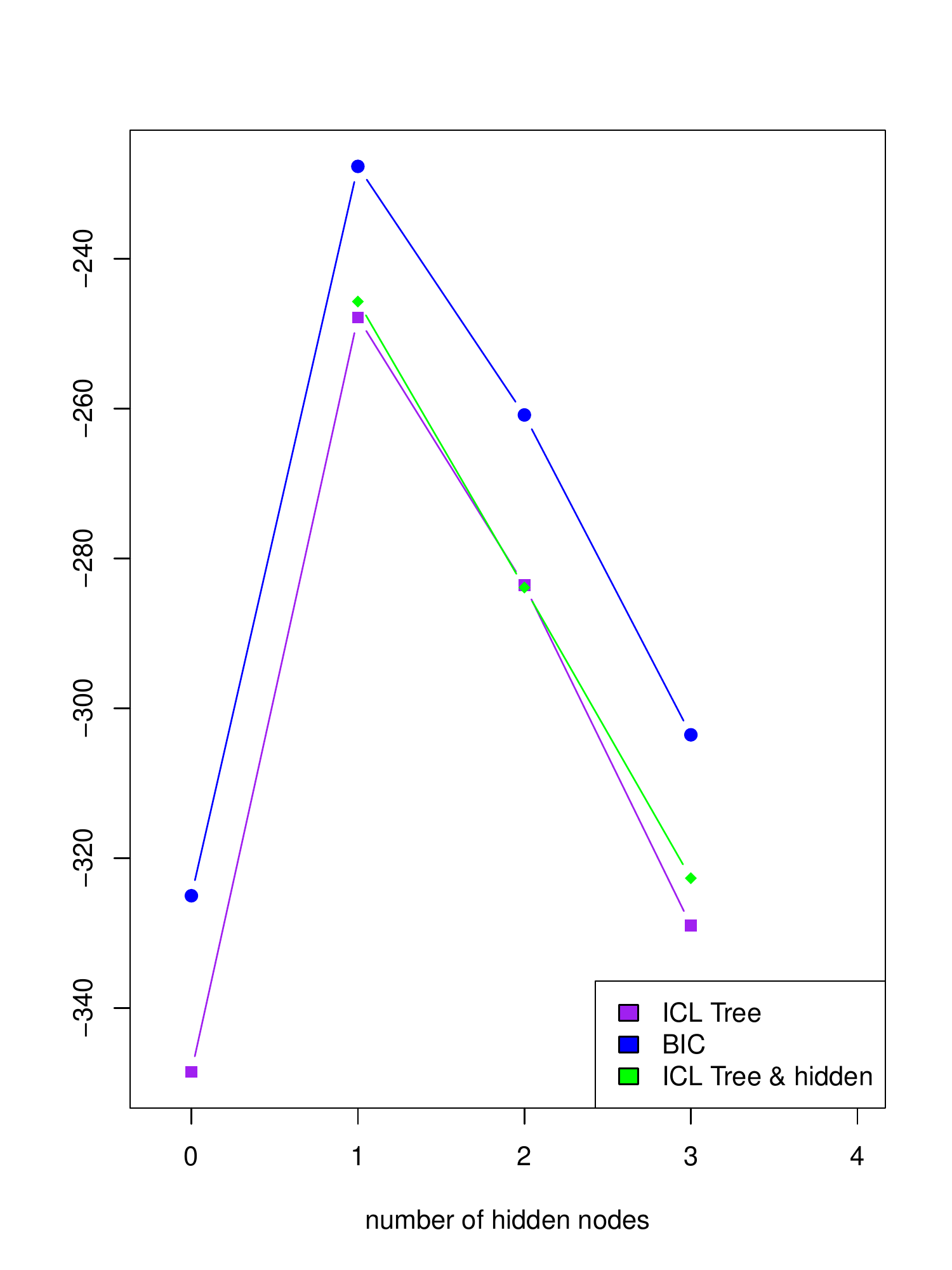} &
  \includegraphics[width=0.4\textwidth, height=0.4\textwidth]{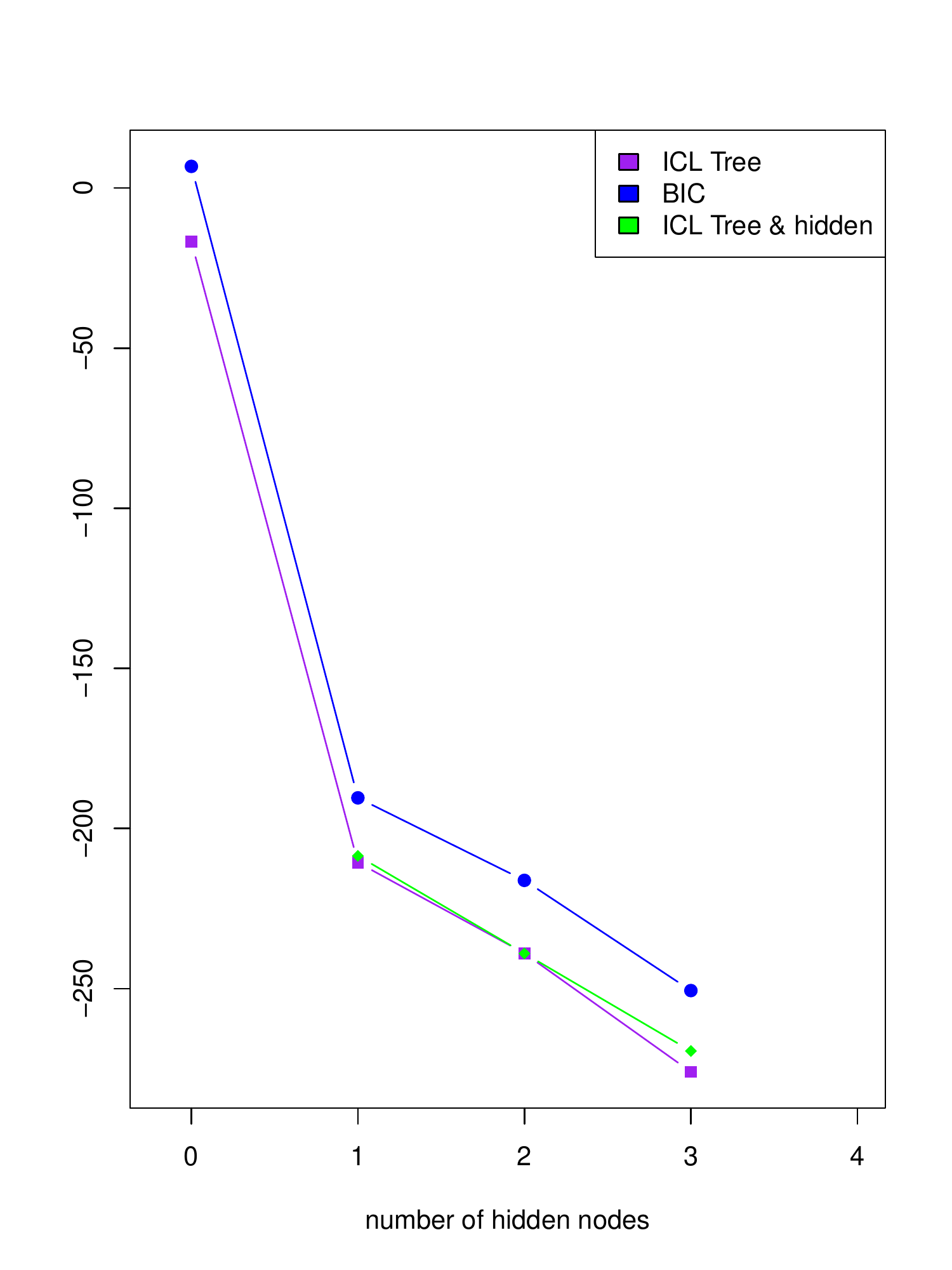} 
\end{tabular}
    \caption{Selection of the number of hidden nodes. Left: when removing one protein. Right: complete dataset.}
    \label{fig:raf-bic}
\end{figure}

Using hierarchical clustering initialization we inferred models with $r = 0$ to $3$ hidden nodes. Figure \ref{fig:raf-bic} (left) shows that the three proposed model selection criteria agree on the true model, that is $r=1$. The same figure shows sthat $ICL_T$ and $ICL_{T, X_H}$ are almost equal and both lower than $BIC$, meaning that the conditional entropy is mostly due to the uncertainty on the tree.

\begin{figure}[!h]
\centering
\begin{tabular}{ccc}
  \includegraphics[width=0.3\textwidth, height=0.3\textwidth]{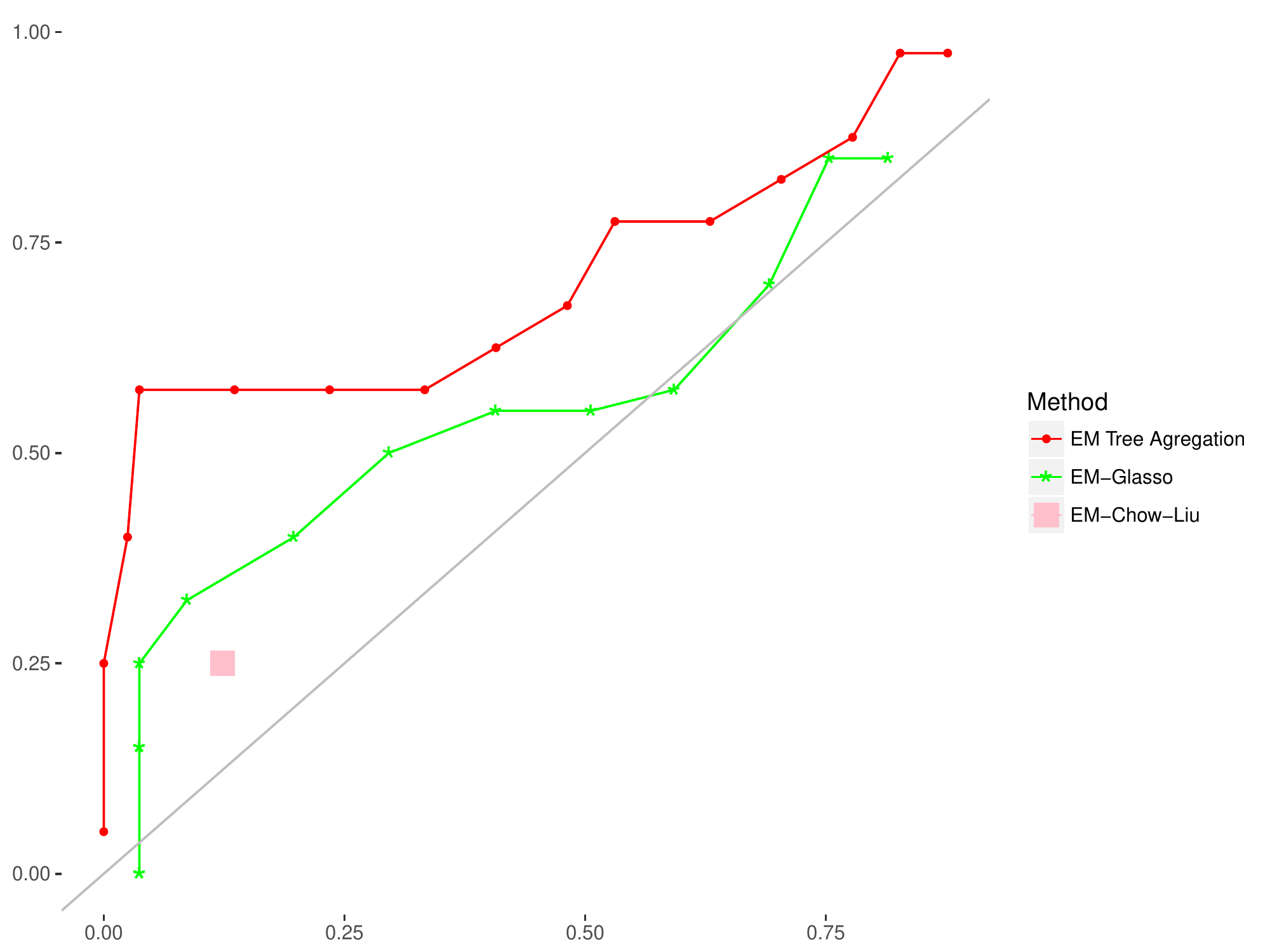} &
  \includegraphics[width=0.3\textwidth, height=0.3\textwidth]{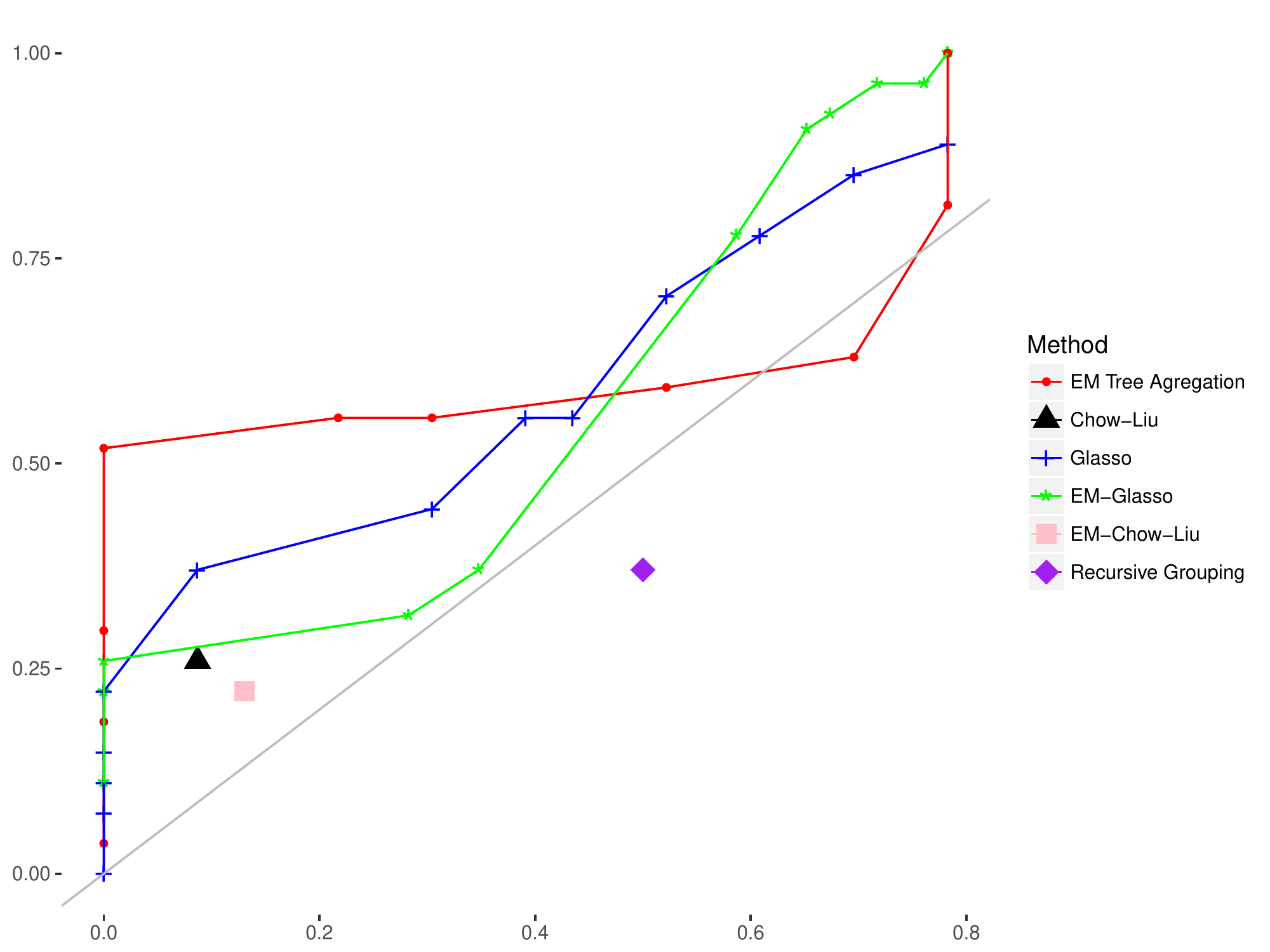} &
  \includegraphics[width=0.3\textwidth, height=0.3\textwidth]{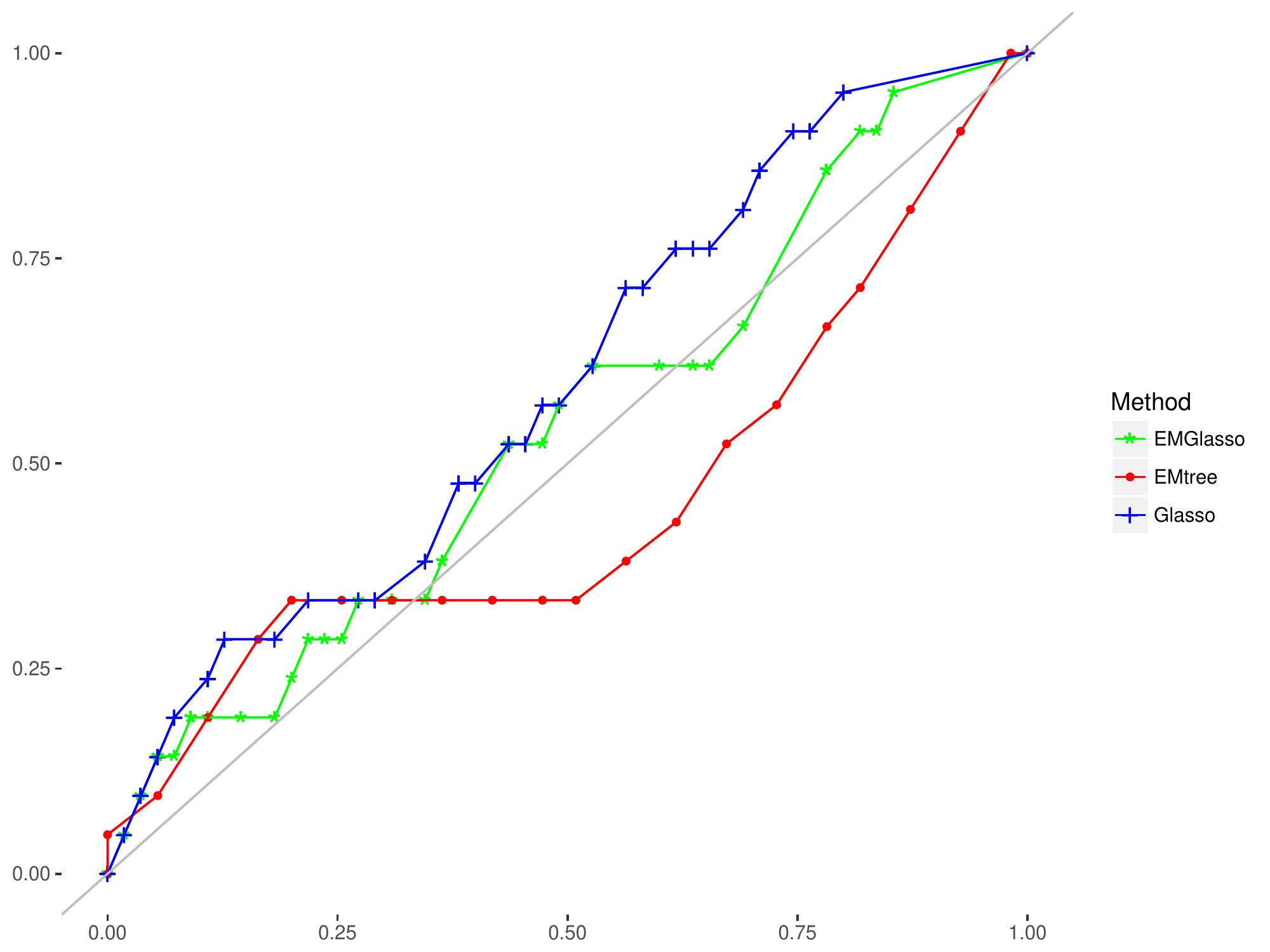} 
\end{tabular}
    \caption{ROC curves for the full (left), marginal (center) graphs and spurious edges (right).}
    \label{fig:raf-roc}
\end{figure}

The performances of the methods described in Section \ref{experiments} are compared on this example in Figure \ref{fig:raf-roc}. The results are similar to those obtained in the simulation study. The proposed latent tree-based approach performs better than the EM-glasso when trying to infer the full graph. The methods also performs well for the marginal graph. In terms of spurious edges, Tree Aggregation displays a plateau, along which the inclusion of spurious edges is delayed compared to Glasso and EM-Glasso.

Finally, we analyzed the complete dataset from \cite{Sachs}, without removing any node. Model selection criteria are given in Figure \ref{fig:raf-bic} (right): they all agree on the absence of a missing node, which is consistent with the biological consensus on the Raf pathway.

\section{Discussion} 

We proposed a method for graphical model inference with missing variables.
Uncovering such a latent structure provides additional hints in the interpretation of the underlying graphical model. For example, the inference of a missing variable allows to pinpoint a group of observed variables, which are related to this unobserved variable. 

Our procedure relies on spanning trees and the computations are performed efficiently using the Matrix-Tree theorem. We have defined a model with a two-layer hidden structure where the graph as
well as the missing nodes are treated as latent variables. We derived conditional distributions of the latent variables given the observations and developed an inference procedure based on the EM algorithm. We also propose model selection criteria to determine the presence
of a hidden structure, as well as the choice of the number of missing variables. We observed on a simulation study that the tree constraint, that we overcome by computing posterior edge probabilities, is not too costly in practice. An implementation of the method is  publicly available through the R package \texttt{LITree}\footnote{The \texttt{LITree} package is available on  GitHub   \url{https://github.com/cambroise/LITree}}. Directions of future work include the extension to non-Gaussian (such as counts) and temporal data.


\bibliographystyle{plainnat}
\bibliography{mybibfile}

\appendix
\section{Computation of the conditional distributions} \label{appendix-1}
We show that the conditional distribution of the tree given the observations factorizes over the edges of the tree.
\begin{equation}
\begin{aligned}
&P(T|X_O) &&\propto P(T)P(X_O|T)\\
& &&\propto\left(\prod_{\{i,j\}\in E_T} \pi_{ij}\right)\underbrace{\frac{\det(K_{T,M})^{\frac{n}{2}}}{(2\pi)^{\frac{np}{2}}}}_{(1)}\underbrace{\exp(-\frac{n}{2}\text{tr}(K_{T,M}\Sigma_O))}_{(2)},
\end{aligned}
\end{equation}
We first focus on the $\det$ term (1). A linear algebra result based on the Schur complement states that
 \begin{equation*}
\begin{aligned}
&\det(K_T)&&=\det\begin{pmatrix}
K_{T,O} & K_{T,OH} \\ 
K_{T,HO} & K_{T,H}
\end{pmatrix}\\
& && =\det(K_{T,H})\det(\underbrace{K_{T,O}-K_{T,OH}(K_{T,H})^{-1}K_{T,HO}}_{K_{T,M}}),
\end{aligned}
\end{equation*}
which finally gives with $\det(K_{T,H})>0$ by definition $\det(K_{T,M})={\det(K_T)} / {\det(K_{T,H})}$. The assumptions on the hidden nodes for identifiability give that $K_{T,H}$ is diagonal and $\det(K_{T,H})=\prod_{h\in H}K_{hh}$ is independent of $T$. Therefore we only need to express $\det(K_T)$ as a product over the edges of $T$. We know from a result of \cite{laurGM} on decomposable graphs that the precision matrix and determinant of tree-structured graphs can be decomposed simply, with $[K_{\{I,J\}}]$ denoting the matrix equal to $K$ on indices $I\times J$ and $0$ elsewhere,
\begin{equation} \label{Eq:defKT}
K_T=\sum_{i\in V}[K_{\{i,i\}}]+\sum_{\substack{\{i,j\}\in V^2}\\\{i,j\}\in E_T}[K_{\{i,j\}}]-[K_{\{i,i\}}]-[K_{\{j,j\}}],
\end{equation}
which gives
\begin{equation}
\text{tr}(K_T\Sigma)=\sum_{i\in V}K_{ii}\Sigma_{ii}+\sum_{\substack{\{i,j\}\in V^2}\\\{i,j\}\in E_T}2K_{ij}\Sigma_{ij}-K_{ii}\Sigma_{ii}-K_{jj}\Sigma_{jj}.
\end{equation}
The approximation mentioned in Section \ref{inference} arises precisely here, where $K_{ii}$ should actually be $K_{T, ii}$. We can also decompose the determinant of $K_T$ as
\begin{equation}
\det(K_T) = \prod_{i\in V} \det([K_{\{i,i\}}]) \prod_{\{i,j\}\in E_T}\frac{\det([K_{\{i,j\}}])}{K_{ii}K_{jj}},
\end{equation}
where $[K_{\{i,j\}}]$ stands for the sub-matrix $K$ where only the $i$th and $j$th rows and columns are kept
and with $\det(K_{T,H}) = \prod_{h\in H} K_{hh}$ and $V=O\bigcup H$,
\begin{equation}
\det(K_{T,M}) = \prod_{i\in O} \det([K_{\{i,i\}}]) \prod_{\{i,j\}\in E_T}\frac{\det([K_{\{i,j\}}])}{K_{ii}K_{jj}}.
\end{equation}

\section{Formulas for the M-step}
\label{appendix-2}

We need to set the derivative of the objective function $E$ given \eqref{eq:ObjFunc} wrt to each $K_{ij}$ to 0. Depending on the status of nodes $i$ and $j$, $K_{ij}$ must satisfy the following:
\begin{align*}
 i, j \in O^2 , i\neq j: \quad & 
 K_{ij}^{h+1}= \left( {1-\sqrt{1+4\widehat{\Sigma}_{ij}^2K_{ii}^hK_{jj}^h}} \right) \left/ {2\widehat{\Sigma}_{ij}} \right.; \\
 i, j \in O\times H : \quad &
 K_{ij}^{h+1}= \left( {-1+\sqrt{1+4(W_{ij}^h)^2K_{ii}^hK_{jj}^h}}\right) \left/ {2W_{ij}^h} \right.; \\
 i = j \in O: \quad &
 \frac{1}{K_{ii}^{h+1}}+\sum_{k\in V}\frac{(K_{ik}^h)^2}{K_{ii}^{h+1}K_{kk}^h-(K_{ik}^h)^2}\alpha^h_{ik} = \widehat{\Sigma}_{ii}; \\
 i = j \in H: \quad &
 \frac{1}{K_{ii}^{h+1}}+\sum_{k\in V}\frac{(K_{ik}^h)^2}{K_{ii}^{h+1}K_{kk}^h-(K_{ik}^h)^2}\alpha^h_{ik} = B^h_{ii}.
\end{align*}

\section{Initialization} \label{app:init}

As the EM-algorithm is highly dependent on its starting point, initialization should be carefully undertaken. As a consequence, although this step is overlooked in most publications, we choose to describe it precisely in this appendix. In our case, it requires an initial graph structure as well as initial values for the missing nodes. Our initialization scheme relies on three stages. First we perform a clustering step and treat the clusters as groups of nodes which share a hidden parent. Then, we initialize the missing variables as the first principal component of the matrix containing their children. Finally, from this completed data, we infer an initial tree using the Chow-Liu algorithm. 

Let us now describe the details of the clustering procedure. We span all the possible triplets of nodes, and merge together the triplet for which the assumption that they had a common hidden parent resulted in the biggest gain in terms of likelihood of the observed realizations. Once the 'best' triplet is selected, we can repeat the same procedure iteratively in order to form clusters in a hierarchical manner. At every level of the hierarchy we have a set of cliques in which the nodes share the same parent and a set of nodes that have not yet been assigned to a clique. For computational reasons we restricted the search to the triplets in which at least one pair of nodes was connected by an edge in the current estimate of the structure. The likelihood gain induced by merging two cliques was penalized for the complexity of the model with the BIC criterion \citep{Sch78}. We show below the dendrogram obtained with this hierarchical clustering procedure, and the cliques (colored nodes) obtained by cutting the hierarchy at the level chosen with BIC. This was done on synthetic data, where we generated $2000$ samples of a Gaussian network with $50$ nodes.

\begin{figure}
\center
\includegraphics[scale=0.3]{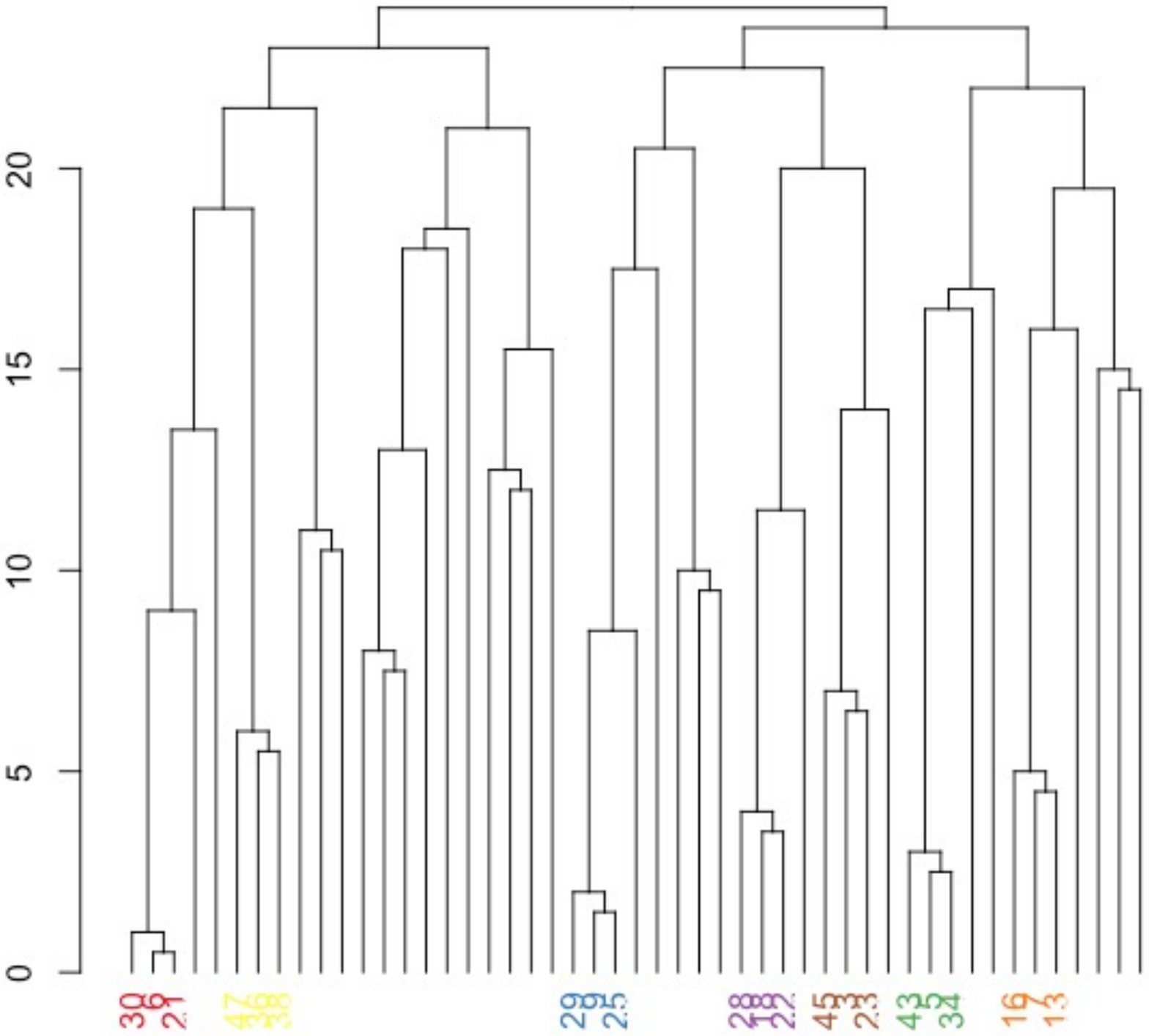}
\caption{Dendrogram of the hierarchical clustering procedure used for initialization. The colored nodes correspond to the clusters at the height chosen with the BIC criterion.}
\end{figure}
\end{document}